\documentclass[aps,prb,
  reprint, %onecolumn,
  tightenlines,
  citeautoscript,%
  showpacs,floatfix,
  superscriptaddress]{revtex4-1}

\usepackage{bm}
\usepackage{amssymb}
\usepackage{graphicx}
\usepackage{amsmath}
\usepackage{textcomp}

\begin{document}
\title{Nonlinear magneto-optic effects in doped graphene and gapped
  graphene: a perturbative treatment} 
\author{J. L. Cheng}
\affiliation{The Guo China-US Photonics Laboratory, Changchun Institute of Optics, Fine Mechanics and
Physics, Chinese Academy of Sciences, 3888 Eastern South Lake Road,
Changchun, Jilin 130033, China.}
\affiliation{University of Chinese Academy of Sciences, Beijing 100049, China}
\author{C. Guo}%
\affiliation{The Guo China-US Photonics Laboratory, Changchun Institute of Optics, Fine Mechanics and
Physics, Chinese Academy of Sciences, 3888 Eastern South Lake Road,
Changchun, Jilin 130033, China.}
\affiliation{The Institute of Optics, University of Rochester, Rochester, NY 14627, USA.}
\date{\today}
\begin{abstract}
  The nonlinear magneto-optic responses are investigated for gapped graphene and doped
  graphene in a perpendicular magnetic field. The electronic states are
  described by Landau levels, and the electron dynamics in an optical field is
  obtained by solving the density matrix in the equation of motion. In the linear dispersion approximation
  around the Dirac points, both linear conductivity and third order nonlinear
  conductivities are numerically evaluated for infrared frequencies. The nonlinear
  phenomena,  including third harmonic generation, Kerr effects and
  two photon absorption, and 
  four wave mixing, are studied. All optical conductivities show
  strong dependence on the magnetic field. At weak magnetic
  fields, our results for doped graphene agree with those in
  the literature. We also present the spectra of the conductivities
  of gapped graphene. At strong  magnetic fields, the
  third order conductivities show peaks with varying the magnetic
  field and the photon energy. These peaks are induced by the resonant transitions between different Landau
  levels. The resonant channels, the positions, and the divergences of
  peaks are analyzed. The conductivities can be greatly modified, up to orders of
  magnitude. The dependence of the conductivities on the gap parameter
  and the chemical potential is studied.   
\end{abstract}
\maketitle

\section{Introduction}
Graphene offers many advantages for applications in photonics
and optoelectronics
\cite{Nat.Photon._4_611_2010_Bonaccorso,ACSNano_6_3677_2012_Bao,ACSNano_8_1086_2014_Low,Nat.Photon._10_227_2016_Sun,Nanoscale_7_4598_2015_Ferrari},
due to tunable optical response and plasmonic excitations in the
mid-infrared to the visible, originating from the gapless linear
dispersion for the low energy excitations. This also leads to its very large optical nonlinearity\cite{Phys.Rep._535_101_2014_Glazov,NewJ.Phys._16_53014_2014_Cheng,Phys.Rev.B_91_235320_2015_Cheng,Phys.Rev.B_93_85403_2016_Mikhailov}, which was
first predicted in theory\cite{Europhys.Lett._79_27002_2007_Mikhailov}
and then demonstrated in
experiments\cite{Phys.Rev.Lett._105_097401_2010_Hendry}. Considering 
the easy integration into silicon based-photonic circuits, it has 
been suggested as an ideal material to provide nonlinear functionality
in photonic devices\cite{Vermeulen_2016}, and the physical origin of 
its optical nonlinearities has attracted a lot of
attentions. Most applications require an efficient way to tune or
control the optical nonlinearities. Recently, it has been proposed that
this can be done by using a
strong perpendicular magnetic field \cite{Phys.Rev.Lett._108_255503_2012_Yao,Phys.Rev.Lett._110_077404_2013_Tokman,J.Phys.Condens.Matter_25_054203_2013_Yao,Phys.Rev.B_86_115427_2012_Rao}.

Due to the linear dispersion, the Landau levels (LLs) of graphene show properties
different from those in conventional two dimensional electron
gas, among which two are especially interesting. One is that the
energies of LLs are not equally spaced, and the energy
difference between adjacent levels decreases with the level
index. The energy of LL at Landau index
$n=\cdots,-2,-1,0,1,2,\cdots$ follows
$\varepsilon_n=\text{sgn}(n)\sqrt{|n|}\hbar\omega_c$ where the cyclotron energy is
$\hbar\omega_c=\sqrt{2\hbar e}v_F \sqrt{B} \approx 36
\sqrt{B(\text{Tesla})}$~meV with the electron charge $-e$ and the Fermi
velocity $v_F=10^6$~m/s. One of the advantages of the nonequidistant
spectrum of LL is that elastic carrier-carrier
scattering can be effectively quenched
\cite{Phys.Rev.B_96_045427_2017_Brem}. The other is that  
the cyclotron energy can be as large as a few tens of meV at several
Tesla for magnetic fields. This suggests the possibility of
applications in the infrared and the LLs of graphene as an excellent
platform for many fundamental physical phenomena, even at room
temperature \cite{Rev.Mod.Phys._83_1193_2011_Goerbig}. Besides many
investigations devoted to the understanding of the linear optical
response in magnetic fields
\cite{J.Phys.Condens.Matter_19_26222_2007_Gusynin,Nat.Phys._7_48_2010_Crassee,Phys.Rev.B_83_075422_2011_Pyatkovskiy},
the study of nonlinear optical effects in LLs 
of graphene starts from the theoretical illustrations of two-color
coherent control of injection currents by Rao and Sipe
\cite{Phys.Rev.B_86_115427_2012_Rao} and of four wave mixing (FWM) by
Yao and Belyanin  
\cite{Phys.Rev.Lett._108_255503_2012_Yao,J.Phys.Condens.Matter_25_054203_2013_Yao}. In
the latter work, a giant bulk effective optical susceptibility 
$\chi^{(3)}_{\text{eff}}\sim 5\times 10^{-9}/B(T)$~m$^2$/V$^2$ was
predicted in full resonant conditions; it was recently
experimentally demonstrated by K\"onig-Otto {\it et al.} in the far
infrared \cite{NanoLett._17_2184_2017_Koenig-Otto}. The use of the strong optical nonlinearity of such systems has been
suggested for generating entangled 
photons \cite{Phys.Rev.Lett._110_077404_2013_Tokman}, for constructing
 all-optical switches \cite{LaserPhys._27_16201_2017_Shiri} and tunable
lasers \cite{NanoLett._17_2184_2017_Koenig-Otto}, for the dynamic
control of coherent pulses \cite{Sci.Rep._7_2513_2017_Yang}, and for
the demonstration of optical bistability and optical multistability
\cite{Phys.B_497_67_2016_Solookinejad,J.Appl.Phys._117_183101_2015_Hamedi}.

{Theoretical treatments in literature include
    Fermi's golden rule \cite{Phys.Rev.B_86_115427_2012_Rao}, dynamics in
    the framework of equation of motion
    \cite{Phys.Rev.Lett._108_255503_2012_Yao,J.Phys.Condens.Matter_25_054203_2013_Yao,Phys.Rev.Lett._110_077404_2013_Tokman,LaserPhys._27_16201_2017_Shiri,Sci.Rep._7_2513_2017_Yang,NanoLett._17_2184_2017_Koenig-Otto,Phys.B_497_67_2016_Solookinejad,J.Appl.Phys._117_183101_2015_Hamedi}, and direct solutions of the Schr\"odinger equation
    \cite{Phys.Rev.B_93_115420_2016_Kibis} by using numerical simulation or by employing the rotating wave
approximation. These studies focus mostly on the transitions between the
lowest few LLs, and  ignore the contributions from other LLs, because
the photon energies are close to the resonant transition
energies. The predicted optical susceptibility $\propto 1/B$ can not be
general at small magnetic field. In addition, there are several other
aspects not well studied in this topic. Firstly, towards a fully understanding 
of the optical nonlinearity for LLs of graphene, it is
necessary to provide a systematic consideration of dependence on the
magnetic field, photon energy, chemical potential and so
on. Standard perturbative calculations  of third order
optical nonlinearities, that usually lead to a preliminary
understanding of the electronic and optical properties, are still
lacking, especially for high photon frequencies. Secondly, with the
increasing interests on the third harmonic generation (THG) of
graphene \cite{arXiv:1710.03694v1,arXiv:1710.04758}, it is also
important to show how the magnetic 
field affects THG responses. Considering the
emergence of many other two
dimensional materials, some of which can be approximately described by a
massive Dirac fermion similar to a gapped graphene, it is of great
interests to understand how the LLs in gapped graphene affect
the optical nonlinearity  
\cite{Phys.Rev.B_92_235307_2015_Cheng,
  Phys.Rev.B_95_035405_2017_Dimitrovski}  
and whether or not the gap can provide an additional level of control,
with  respect to opening a gap in graphene for further electronic and
photonic applications. Lastly, the light-matter interaction used in
published works is mostly described in  the velocity gauge (the
$\bm p\cdot\bm A$ interaction).
Without any approximation, the velocity gauge is equivalent to the length
gauge (the $\bm r\cdot\bm E$ interaction) for homogeneous
fields. However, this 
equivalence may be broken with adopting approximations of
      truncated bands and finite region in the Brillouin zone; then
      the calculation of optical response in the velocity gauge
      requires a very careful treatment due to the appearance of
      unphysical ``false'' divergences
      \cite{Phys.Rev.B_48_11705_1993_Sipe}, which can be fixed by
      additional efforts of employing the sum rules or conservation
      laws. Besides, when the linear dispersion approximation is used for
      graphene, Wang {\it et al.} \cite{Phys.Rev.B_94_195442_2016_Wang}
      identified and solved a different divergent problem for the linear
      conductivity, where the integration over wave vector become
      divergent for all photon energies. It is widely accepted that the
      calculation in the length gauge can avoid all these
      problems without any additional effort. Although it is not
      clear what problem might be induced in discrete level systems
      from the velocity gauge,  benchmark calculations in the length
      gauge \cite{Phys.Rev.B_52_14636_1995_Aversa}, which does not
      lead to such difficulties, would be helpful.  
}

In this work we present  perturbative expressions for the linear
conductivity and third order conductivity of gapped
graphene (GG) and doped graphene (DG) subject to a perpendicular
magnetic field, where the light matter interaction
is treated in the length gauge $\bm r\cdot\bm E$ to avoid the use of
sum rules and conservation laws. We focus on the spectrum of the linear
conductivity and those of the third order 
conductivities for different nonlinear phenomena, including {THG}, Kerr effects and two photon absorption (or
nonlinear corrections to the linear conductivity, NL), and FWM. 
We consider the limit as the magnetic field goes to 
zero, and compare the conductivities of doped graphene with those
obtained without the presence of the magnetic field. Furthermore, we
present the nonlinear conductivities of gapped graphene, which has
only been the subject of a few studies. At a strong magnetic field,
we show resonances between discrete LLs and identify the
condition for the resonances to arise.  

We organize this paper as follows: In section~\ref{sec:model} we present
a model Hamiltonian, the matrix elements of the optical
dipoles, the equation of motion involving external
optical fields, and the perturbative expressions for optical
conductivities. In section~\ref{sec:weakfield}, we discuss the limits
of the optical conductivity at weak magnetic field, and compare with
the well-known conductivities at zero magnetic field. In the same framework we also present the conductivity
for gapped graphene. In section~\ref{sec:strongfield}, we consider
the magnetic field 
dependence of the optical conductivities and discuss the conditions for
resonant transitions.  In section~\ref{sec:spectraStrongB}, we show the
spectra of the optical conductivities at a strong magnetic field. We
conclude in section~\ref{sec:conclusion}.

\section{Model\label{sec:model}}
Under a perpendicular magnetic field $\bm B=B \hat{\bm z}$, the
electronic states around the Dirac points  of graphene are determined
by an effective Hamiltonian \cite{Phys.Rev.Lett._108_255503_2012_Yao}
\begin{equation}
  H^0 = \begin{pmatrix} H_{+;\bm p+e\bm A(\bm r)} & 0 \\ 0 & H_{-;\bm
      p+e\bm A(\bm r)} 
  \end{pmatrix}
\end{equation}
with taking the electron charge as $-e$, the vector potential
\begin{equation}
  \bm A(\bm r) = B x \hat{\bm y}\,,\label{eq:vecpot}
\end{equation} 
and the Hamiltonian in each valley
\begin{equation}
  H_{\nu;\bm p}^0 =  v_F(p_x \sigma_x + \nu p_y \sigma_y) + \Delta
  \sigma_z\,. \label{eq:modelh}
\end{equation}
Here $\nu$ is the valley index taking a value $\nu=+$ for the $\bm K$
valley or $\nu=-$ for the $\bm K^{\prime}$ valley, $v_F$ is the Fermi velocity,
$\Delta$ is a mass parameter to induce an energy gap $2\Delta$ in the
absence of a magnetic field, and $\sigma_i$ ($i=x,y,z$) are the Pauli
matrices. The mass parameter, corresponding to
asymmetric on-site energies, could be induced by a Si-terminated SiC
substrate\cite{Nat.Mater._6_770_2007_Zhou} or a BN substrate. We use
$H^0(B,\Delta)$ to explicitly show the magnetic field and mass
parameter dependence. By using the transformation 
\begin{eqnarray*}
  H^0(-B,-\Delta) &=& {\cal T}^{-1} H^0(B,\Delta) {\cal T}\,,\\
  H^0(B,-\Delta) &=& \begin{pmatrix} 0 & \sigma_x \\ \sigma_x & 0
  \end{pmatrix} H^0(B,\Delta) \begin{pmatrix} 0 & \sigma_x \\ \sigma_x & 0
  \end{pmatrix}\,,
\end{eqnarray*}
where ${\cal T}=i\sigma_yK$ is similar to a time reversal operator and  $K$ is the complex conjugation operator, we only need to
discuss the parameter domain $B\ge0$ and $\Delta\ge0$. 

Obviously, the $\bm K$ and $\bm K^\prime$ valleys are not coupled in
this model, but they are connected through 
\begin{equation}
  H_{+;\bm p} = -\sigma_y H_{-;\bm p} \sigma_y\,.\label{eq:valleyeq}
\end{equation}
Thus we can obtain the electronic states in the $\bm K^\prime$ valley
from those in the $\bm K$ valley by
utilizing this transformation.

\subsection{Eigenstates and eigen energies}
We first solve the electronic states in the $\bm K$ valley for the
parameter domain $B\ge0$ and $\Delta\ge0$. 
For the chosen vector potential in Eq.~(\ref{eq:vecpot}), there exists
translation symmetry along the $y$ direction, thus the eigen states
can be written as
$\Psi(\bm r) = \frac{1}{\sqrt{2\pi}} e^{ik y} \Phi(x+l_c^2k)$.
Here $k$ is a quasi wave vector along the $y$ direction,
$l_c=\sqrt{\hbar/(eB)}$ is the magnetic length, and $\Phi(x)$ is a spinor
envelope wave function, which satisfies 
\begin{equation}    
   \begin{pmatrix} \Delta & v_F(p_x -ieBx) \\
     v_F(p_x+ieBx) & -\Delta
   \end{pmatrix}\Phi(x) = E \Phi(x)\label{eq:ll0}
\end{equation}
This eigen equation is solved by employing creation and
annihilation operators for LLs
\begin{equation}
  \hat a = \frac{l_c}{\sqrt{2}\hbar} (p_x - i e B x) \,,\quad
  \hat a^{\dag} =\frac{l_c}{\sqrt{2}\hbar} (p_x + i e B x)\,,
\end{equation}
with $  [\hat a,\hat a^{\dag}] = 1$. 
The eigenstates of the particle number operator $\hat a^{\dag}\hat
a$ are harmonic oscillator states, $\phi_{n}(x)$ $(n=0,1,2,\cdots)$,
which are determined from
\begin{eqnarray}
  \hat a \phi_0(x) &=&  0\,,\quad \Longrightarrow\quad \phi_0(x) =\frac{1}{(\sqrt{\pi}l_c)^{1/2}} \exp\left(-\frac{x^2}{2l_c^2}\right)\notag\,,\\
  \hat a^{\dag}\phi_n(x) &=& \sqrt{n+1}\phi_{n+1}(x)\,,\quad  \hat a\phi_n(x) = \sqrt{n} \phi_{n-1}(x)\,.\label{eq:operator}
\end{eqnarray}
Then Eq.~(\ref{eq:ll0}) becomes 
\begin{equation*}
  \begin{pmatrix} \Delta & \hbar\omega_c\hat a \\ \hbar\omega_c \hat a^{\dag} & -\Delta
  \end{pmatrix} \Phi(x) = E\Phi(x)\,,
\end{equation*}
with the cyclotron energy $\hbar\omega_c=\sqrt{2}\hbar v_F/l_c$. By
expanding the eigenstates in the basis of
$\{\phi_n(x), n=0,1,2,\cdots\}$,
$\Phi(x) = \sum_{n}\begin{pmatrix}\varphi_{1;n} \\ \varphi_{2;n} \end{pmatrix} \phi_{n}(x)$, all eigenstates and eigen
energies can be identified as
\begin{eqnarray}
  \varepsilon_{sn} &=& s \epsilon_n\,,\quad \epsilon_n=\sqrt{\Delta^2 + n(\hbar\omega_c)^2}\,,\\
  \Phi_{sn}(x) &=& \frac{1}{\sqrt{2}} \begin{pmatrix}
    s \sqrt{1+s\alpha_n} \phi_{n-1}(x) \\ \sqrt{1-s\alpha_n} \phi_{n}(x)
  \end{pmatrix}\,.
\end{eqnarray}
Here $s$ is a band index taking a value from $+$ for the upper band or
$-$ for the lower band; $n$ is a Landau index
for LLs taking a value from $n=0,1,2,\cdots$ for $s=-$ or
$n=1,2,\cdots$ for $s=+$; and 
$\alpha_n=\Delta/\epsilon_n$. We have
$\alpha_0=1$ for $\Delta>0$, and we use the convention
$\phi_{-1}(x)\equiv0$. The level $n=0$ is special. 

We use an additional band index $s$ to label the LLs, which is
different from the conventional label used in the literature
\cite{Phys.Rev.Lett._108_255503_2012_Yao}. However, our notation
provides an easy classification of the optical transitions, including intraband transitions, occurring   
inside one band, and interband transitions, occurring between these two 
bands. This is very useful for understanding the results at weak
magnetic fields. In the limit of $B\to0$, the
energy of LLs becomes continuous, but they can still be treated as an
orthogonal and complete basis. 
\begin{figure}[tph]
  \centering
  \includegraphics[width=7.5cm]{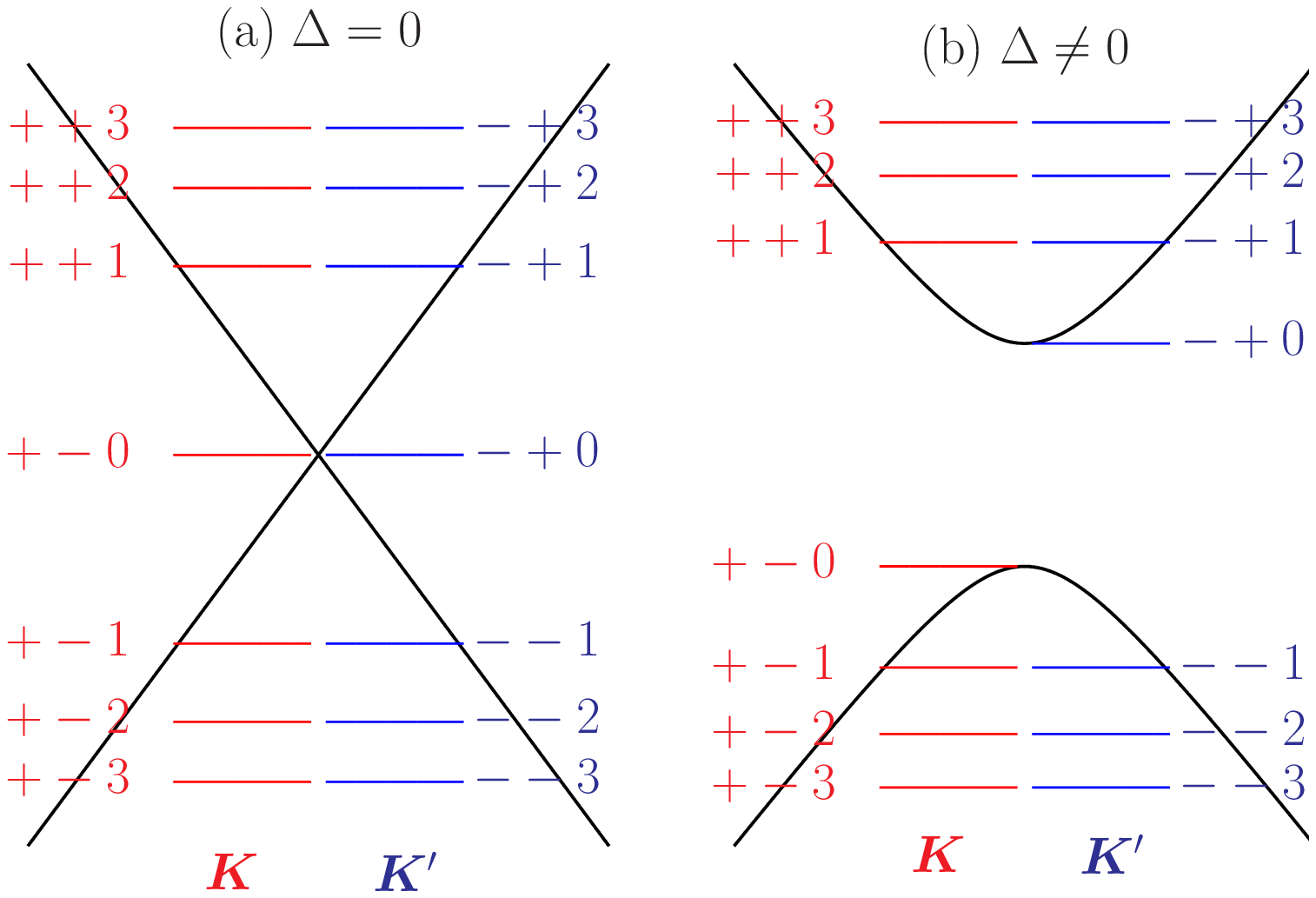}\\
  \includegraphics[width=7.5cm]{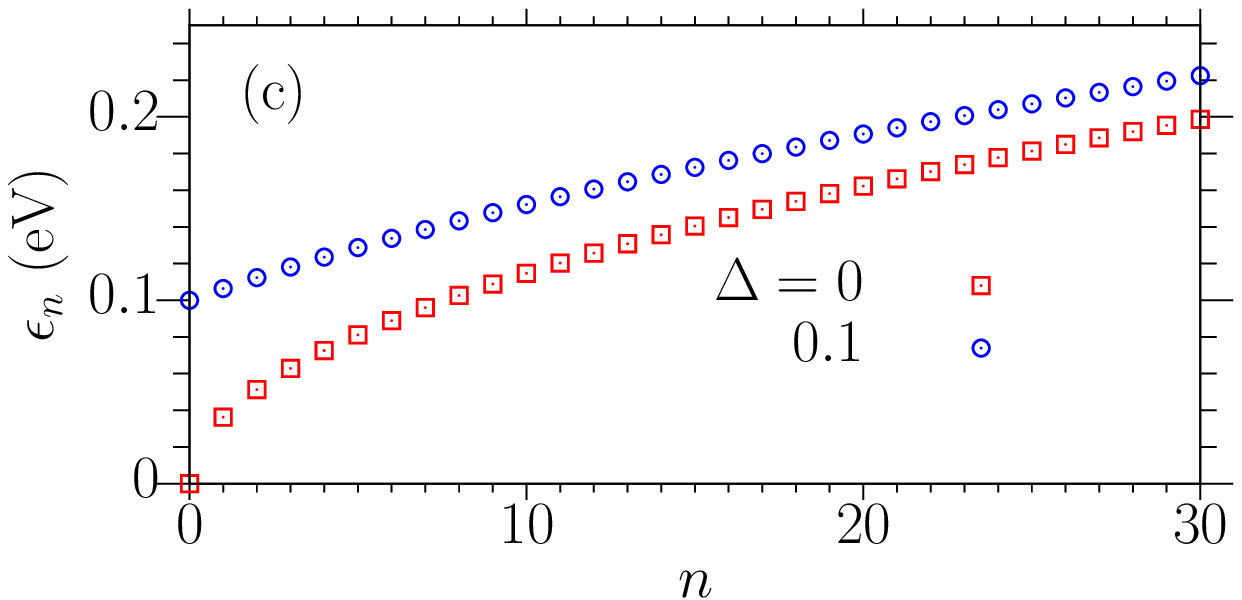}
\caption{(color online) Illustrations of Landau levels in both valleys
  for (a) $\Delta=0$ and (b) $\Delta\neq 0$. The black curves are the band structures
  for zero magnetic field. The Landau levels in $\bm K$ ($\bm
  K^{\prime}$) valley are given as red (blue) horizontal lines with
  level indices $\nu s n$ aside. (c) The energies of Landau levels for
  (red square) $\Delta=0$ and (blue circle) $\Delta=0.1$~eV.}
\label{fig:illus}
\end{figure}

To summarize, all eigenstates  and eigen energies in the $\bm K$
valley can be labeled by indices $\{\nu s n k\}$ with $\nu=+$, and are given as 
\begin{eqnarray}
  \Psi_{+snk}(\bm r)&=&\frac{1}{2\sqrt{\pi}} e^{iky}\begin{pmatrix}s \sqrt{1+s\alpha_n} \phi_{n-1}(x+l_c^2k) \\ \sqrt{1-s\alpha_n} \phi_{n}(x+l_c^2k)
  \end{pmatrix}\,,\quad\\
  E_{+snk} &=& s \epsilon_n \,.
\end{eqnarray}
Using Eq.~(\ref{eq:valleyeq}), we can obtain all eigenstates
and eigen energies in the $\bm K^{\prime}$ valley as
\begin{eqnarray}
  \Psi_{-snk}(\bm r)& = & \sigma_y \Psi_{+\bar{s}nk}(\bm r)\,,\\
  E_{-snk}& = & -E_{+\bar{s}nk}\,,
\end{eqnarray}
where the band index $s$ is reversed by hand. They can be
organized into a unified form as 
\begin{equation}
  {  \renewcommand{\arraystretch}{1.8}
    \begin{array}{rcl}
      E_{\nu snk} &=& s \epsilon_n\,, \quad \Psi_{\nu s nk}(\bm r) = \frac{1}{\sqrt{2\pi}} e^{iky} \Phi_{\nu s n}(x+l_c^2k)\,,\\
      \Phi_{\nu s n }(x) &=&\frac{1}{\sqrt{2}} \begin{pmatrix} s \sqrt{1+ s \alpha_n} \phi_{n-(\nu+1)/2}(x) \\ \sqrt{1- s \alpha_n} \phi_{n + (\nu - 1)/2}(x)
      \end{pmatrix}\,,\\
      \nu &=& \pm,\quad s = \pm\,,\notag\\
      n &\in& \text{integers and } n\ge \frac{1+\nu s}{2}\,.
  \end{array}}
\end{equation}
The parameter space
of $n$ depends on both $s$ and $\nu$. 
The eigen energy $E_{\nu s n k}$  depends only on the band index $s$
and Landau level index $n$, and each of them has a degeneracy induced
by $k$ as ${\cal
  D}=\frac{1}{2\pi l_c^2}g_s$ with 
$g_s=2$ for spin degeneracy.\footnote{Simply, we can consider a sample with
  dimensions $L_x\times L_y$. The interval of $k$ should be
  $\frac{2\pi}{L_y}$, and in the region $0<l_c^2k<L_x$ the number of
  the states is 
  $\frac{L_x}{l_c^2}/\left(\frac{2\pi}{L_y}\right)=\frac{L_xL_y}{2\pi
    l_c^2}$, which gives the density of states without spin degeneracy
  as $\frac{1}{2\pi l_c^2}$.} 
Figure~\ref{fig:illus} illustrates
the LLs in both valleys for $\Delta=0$  and
$\Delta=0.1$~eV. For nonzero $\Delta$, the Landau level $n=0$
in the $\bm K$ valley is at the top of the lower band, while the one in
the $\bm K^{\prime}$ valley is at the bottom of the upper band; As
$\Delta$ goes to zero, both of them are at the Dirac points. 
In this work, we take the Fermi velocity as $v_F=10^6$~m/s, then the
cyclotron energy is $\hbar\omega_c\approx
36.3\sqrt{B/1~\text{T}}$~meV. In Fig.~\ref{fig:illus} (c), we show the
energies of $\epsilon_n$ for $\Delta=0$ and
$0.1$~eV. 

\subsection{Equation of motion and perturbative optical conductivity}
When a uniform electric field $\bm E(t)$ is applied, the total
Hamiltonian is 
\begin{equation}
  H_{\nu} = H_{\nu;p+ e\bm A(\bm r)}^0 + e\bm E(t)\cdot\bm r\,.
\end{equation}
In the second quantization form, the total Hamiltonian becomes
\begin{equation}
  \hat H(t)=\sum_{\nu sn} s\epsilon_n \int dk \hat c_{\nu
    snk}^{\dag}(t)\hat c_{\nu snk}(t)+ e\bm
  E(t)\cdot \hat{\bm r}(t) \,, \label{eq:H}
\end{equation}
Here $\hat c_{\nu snk}(t)$ is the annihilation operator in Heisenberg
picture for the state $\Psi_{\nu snk}(\bm r)$, and it satisfies the anti-commutator
$\{\hat c_{\nu_1 s_1n_1k_1},\hat c_{\nu_2 s_2n_2k_2}^{\dag}\}=\delta_{\nu_1,\nu_2}\delta_{s_1,s_2}\delta_{n_1,n_2}\delta(k_1-k_2)$
and $\{\hat c_{\nu_1 s_1n_1k_1},\hat c_{\nu_2 s_2n_2k_2}\}=0$;
$\hat{\bm r}(t)$ is the position operator in the second quantization form,
which is given by
\begin{eqnarray}
\hat{\bm r}(t)  &=& \sum_{\nu s_1s_2\atop n_1n_2}\bm
  \xi_{\nu;s_1n_1,s_2n_2}\int dk \hat c_{\nu s_1n_1k}^{\dag}(t) \hat c_{\nu s_2n_2k}(t)
  \notag\\
  &&+\sum_{\nu sn} \int dk \hat c_{\nu snk}^{\dag}(t)\left(-l_c^2 k \hat{\bm
    x}-i\hat{\bm y}\frac{\partial}{\partial k} \right)\hat c_{\nu snk}(t)\,.\label{eq:opr}\quad\quad
\end{eqnarray}
The term $\bm \xi_{\nu;s_1n_1,s_2n_2}$ is the Berry connection between
states $\Psi_{\nu s_1n_1}(\bm r)$ and $\Psi_{\nu s_2n_2}(\bm r)$. In
the presence of magnetic field, it is convenient to calculate the
circularly polarized components  $\xi^\delta$ defined from a vector
$\bm\xi = \sum_{\tau=\pm} \xi^\tau \hat{\bm e}^{\overline{\tau}}$ with
$\xi^\pm =\frac{1}{\sqrt{2}}(\xi^x\pm i \xi^y)$ and
$\hat{\bm e}^\pm =\frac{1}{\sqrt{2}}(\hat{\bm x} \pm i \hat{\bm y})$.
Here $\tau$ is an index for the circularly
polarization direction with taking value from $\pm$, and the notation
$\overline{\tau}$ means $\overline{\tau}=-$ ($+$) for $\tau=+$ ($-$).
The calculation of Berry connections is listed in
Appendix~\ref{app:v}, which gives
\begin{eqnarray}
  \xi_{\nu;s_1n_1,s_2n_2}^\tau &=& -i\tau  w_{s_1(n_2+\tau),s_2n_2}^{(\nu\tau)} \delta_{n_1,n_2+\tau}\,,\label{eq:bc}\\
  w_{s_1n_1,s_2n_2}^{(+)} &=& \frac{l_c}{2}\left(
  s_1s_2\sqrt{1+s_1\alpha_{n_1}}\sqrt{1+s_2\alpha_{n_2}}
  \sqrt{n_2}\right.\notag\\
  &&\left.+
  \sqrt{1-s_1\alpha_{n_1}}\sqrt{1-s_2\alpha_{n_2}}
  \sqrt{n_1}\right)\,,\notag\\
  w_{s_1n_1,s_2n_2}^{(-)} &=& w_{s_2n_2,s_1n_1}^{(+)}\,.\notag
\end{eqnarray}
Here $(\nu\tau)=(+)$ for $\nu=\tau$ and $(-)$ for
$\nu=\overline{\tau}$.  
The selection rules for $\xi^\tau_{\nu;s_1n_1,s_2n_2}$ between
different LLs $n_1$ and $n_2$ are $n_1=n_2+\tau$, which is
independent of the band index.  

The velocity operator is $\hat{\bm v}(t)=[\hat{\bm r}(t), \hat
  H(t)]/(i\hbar)$, and the current density operator  $\hat
{\bm J}(t)=-e\hat{\bm v}(t)$ is 
\begin{equation}
  \hat{\bm J}(t) =-e \sum_{\nu s_1s_2\atop n_1n_2}\bm v_{\nu;s_1n_1,s_2n_2}
  \int dk \hat c_{\nu s_1n_1k}^{\dag}(t)\hat c_{\nu s_2n_2k}(t)\,.
\end{equation}
with the matrix elements
\begin{equation*}
\bm v_{\nu;s_1n_1,s_2n_2} = i\hbar^{-1}(s_1\epsilon_{n_1}-s_2\epsilon_{n_2}) \bm \xi_{\nu;s_1n_1,s_2n_2}\,.
\end{equation*}

Because the eigen energies and the velocity matrix elements do not
depend on $k$, {the dynamics of the current density relies on the dynamics of the
  effective density matrix operator $\hat\rho_{\nu}(t)$,
  which is defined as
  \begin{equation}
    \hat\rho_{\nu;s_1n_1s_2n_2}(t)={\cal D}^{-1}\int dk \hat
  c_{\nu;s_2n_2k}^{\dag}(t) \hat c_{\nu;s_1n_1k}(t)\,,
  \end{equation}
  with ${\cal D}=1/(\pi l_c^2)$ being the LL degeneracy. The time evolution of
  $\hat\rho_\nu(t)$ is determined by the Heisenberg equation 
\begin{equation}
  i\hbar\frac{\partial\hat \rho_{\nu}(t)}{\partial t} = [H_\nu, \hat\rho_\nu(t)] +
  i\hbar\left.\frac{\partial\hat \rho_\nu(t)}{\partial t}\right|_{scat}\,,
\end{equation}
where the last term $\left.\frac{\partial \hat\rho_\nu(t)}{\partial
  t}\right|_{scat}$ is the scattering term. By taking the expectation
value on both side of above equation with respect to the equilibrium
state, we get the matrix elements of 
$\rho_{\nu;s_1n_1s_2n_2}(t)=\langle \hat \rho_{\nu;s_1n_1s_2n_2}(t)
\rangle$ satisfying the equation of motion
}
\begin{eqnarray}
  i\hbar \frac{\partial \rho_{\nu;s_1n_1,s_2n_2}(t)}{\partial t} &=&
  (s_1\epsilon_{n_1}-s_2\epsilon_{n_2}) \rho_{\nu;s_1n_1,s_2n_2}(t) \notag\\
  &+&  e\bm
  E(t)\cdot \sum_{sn} \left[
    \bm\xi_{\nu;s_1n_1,sn}\rho_{\nu;sn,s_2n_2}(t)\notag\right.\\
    && \left.- \rho_{\nu;s_1n_1,sn}(t)
    \bm \xi_{\nu;sn,s_2n_2}\right]\notag\\
  &-&  i\Gamma\left[\rho_{\nu;s_1n_1,s_2n_2}(t) - \rho_{\nu;s_1n_1,s_2n_2}^0\right]\,.\quad\quad\label{eq:ksbe}
\end{eqnarray}
Here we describe the scattering by phenomenological relaxation
processes with only one relaxation parameter $\Gamma$,
$\rho_{\nu;s_1n_1,s_2n_2}^0=\delta_{s_1,s_2}\delta_{n_1,n_2}f_{s_1n_1}$
is the density matrix element at the equilibrium,
$f_{sn}=[1+e^{(s\epsilon_{n}-\mu)/(k_BT)}]^{-1}$ is the Fermi-Dirac
distribution with a chemical potential $\mu$ and temperature 
$T$. {Note that the $k$ derivative appearing in the Hamiltonian $H(t)$
  in Eqs.~(\ref{eq:H}) and (\ref{eq:opr}) does not contribute to the equation of motion, because
    both the current operator and the density matrix are only related
    to a term like $\int dk \hat{c}_k^\dag\hat{c}_k$.}

After some algebra listed in the Appendix.~\ref{app:j}, we get the
perturbative linear and third order conductivities.
Using the selection rules of $\bm\xi$ and $\bm v$, the condition for
nonzero components of $\sigma^{(1);\tau\alpha}$ is $\tau=\alpha$,
and that of $\widetilde{\sigma}^{(3);\tau\alpha\beta\gamma}$ is
$\tau=\alpha+\beta+\gamma$. Thus the possible nonzero components of
third order conductivity are $\widetilde{\sigma}^{(3);\tau\tau\tau\overline{\tau}}$,
$\widetilde{\sigma}^{(3);\tau\tau\overline{\tau}\tau}$, and
$\widetilde{\sigma}^{(3);\tau\overline{\tau}\tau\tau}$. The linear
conductivity is expressed as
\begin{widetext}
\begin{eqnarray}
  \sigma^{(1);\tau\tau}(\omega) &=& -\frac{ie^2}{\hbar}{\cal D}
  \sum_{\nu s_1 s_2}\sum_{n} \frac{(s_2\epsilon_{n +
      \tau}-s_1\epsilon_{n})[w_{s_2(n+\tau),s_1n}^{(\nu\tau)}]^2
    (f_{s_2(n+\tau)}-f_{s_1n})}{\hbar\omega+i\Gamma-(s_1\epsilon_{n}-s_2\epsilon_{n+\tau})}\,, \label{eq:sigma1}
\end{eqnarray}
Although the inversion symmetry is broken for a nonzero mass term
$\Delta$, the second order response of optical current is still zero
in our  approach because the linear dispersion approximation in
$H_{\nu;\bm p}^0$ includes an additional inversion symmetry; the
nonzero second order response can be obtained beyond the linear
dispersion approximation. The third order conductivities are
  {\allowdisplaybreaks
\begin{eqnarray}
  \sigma^{(3);\tau\alpha\beta\gamma}(\omega_1,\omega_2,\omega_3)
  &=&\frac{1}{6}
  \left[\widetilde{\sigma}^{(3);\tau\alpha\beta\gamma}(w,\hbar(\omega_2+\omega_3)+i\Gamma,\hbar\omega_3+i\Gamma)
    +
    \widetilde{\sigma}^{(3);\tau\alpha\gamma\beta}(w,\hbar(\omega_2+\omega_3)+i\Gamma,\hbar\omega_2+i\Gamma)\right.\notag\\
&&+\widetilde{\sigma}^{(3);\tau\beta\alpha\gamma}(w,\hbar(\omega_1+\omega_3)+i\Gamma,\hbar\omega_3+i\Gamma)
    +
    \widetilde{\sigma}^{(3);\tau\beta\gamma\alpha}(w,\hbar(\omega_1+\omega_3)+i\Gamma,\hbar\omega_1+i\Gamma)\notag\\
  &&\left.+\widetilde{\sigma}^{(3);\tau\gamma\alpha\beta}(w,\hbar(\omega_1+\omega_2)+i\Gamma,\hbar\omega_2+i\Gamma)
    +
    \widetilde{\sigma}^{(3);\tau\gamma\beta\alpha}(w,\hbar(\omega_1+\omega_2)+i\Gamma,\hbar\omega_1+i\Gamma)    \right]\,,
\end{eqnarray}
with $w=\hbar(\omega_1+\omega_2+\omega_3)+i\Gamma$ and
\begin{eqnarray}
  && \widetilde{\sigma}^{(3);\tau\alpha\beta\gamma}(w,w_0,w_3)\notag\\
  &=& -\frac{ie^4}{\hbar}{\cal D} \sum_{\nu s_1s_2\atop
     s_3s_4}\sum_{n}\frac{(s_2\epsilon_{n}-s_1\epsilon_{n-\tau})
     w_{s_2n,s_1(n-\tau)}^{(\nu\tau)}}{w-(s_1\epsilon_{n-\tau}-s_2\epsilon_{n})}\frac{
     w_{s_1(n-\tau),s_3(n-\gamma-\beta)}^{(\nu\overline{\alpha})}}{w_0-(s_3\epsilon_{n-\gamma-\beta}-s_2\epsilon_{n})}\notag\\
   &&\times \left[\frac{
      w_{s_3(n-\gamma-\beta),s_4(n-\gamma)}^{(\nu\overline{\beta})}
      w_{s_4(n-\gamma),s_2n}^{(\nu\overline{\gamma})}
      (f_{s_2n}-f_{s_4(n-\gamma)})
    }{w_3-(s_4\epsilon_{n-\gamma}-s_2\epsilon_{n})}\right.\notag\\
    &&\quad\quad \left.-
    \frac{w_{s_3(n-\beta-\gamma),s_4(n-\beta)}^{(\nu\overline{\gamma})}w_{s_4(n-\beta),s_2n}^{(\nu\overline{\beta})}
      (f_{s_4(n-\beta)}-f_{s_3(n-\beta-\gamma)})
    }{w_3-(s_3\epsilon_{n-\beta-\gamma}-s_4\epsilon_{n-\beta})}\right]
  \notag\\
  &+& \frac{ie^4}{\hbar}{\cal D}\sum_{\nu s_1s_2\atop s_3 s_4} \sum_n\frac{(s_2\epsilon_{n}-s_1\epsilon_{n-\tau})
     w_{s_2n,s_1(n-\tau)}^{(\nu\tau)}}{w-(s_1\epsilon_{n-\tau}-s_2\epsilon_{n})}\frac{
    w_{s_4(n-\alpha),s_2n}^{(\nu\overline{\alpha})}}{w_0-(s_1\epsilon_{n-\tau}-s_4\epsilon_{n-\alpha})}
  \notag\\
  &&\times \left[
    \frac{w_{s_1(n-\tau),s_3(n-\gamma-\alpha)}^{(\nu\overline{\beta})}
      w_{s_3(n-\gamma-\alpha),s_4(n-\alpha)}^{(\nu\overline{\gamma})}
      (f_{s_4(n-\alpha)}-f_{s_3(n-\gamma-\alpha)})
    }{w_3-(s_3\epsilon_{n-\gamma-\alpha}-s_4\epsilon_{n-\alpha})}
    \right.\notag\\
    && \quad\quad \left.- \frac{w_{s_1(n-\tau),s_3(n-\alpha-\beta)}^{(\nu\overline{\gamma})} w_{s_3(n-\alpha-\beta),s_4(n-\alpha)}^{(\nu\overline{\beta})} (f_{s_3(n-\alpha-\beta)}-f_{s_1(n-\tau)}) }{w_3-(s_1\epsilon_{n-\tau}-s_3\epsilon_{n-\alpha-\beta})}
    \right]\,.\label{eq:sigma3}
\end{eqnarray}
}\end{widetext}
It is constructive to give the tensor components in Cartesian
coordinates. The independent Cartesian components of the linear
conductivity are $\sigma^{(1);xx}(\omega)=\sigma^{(1);yy}(\omega)$ and
$\sigma^{(1);yx}(\omega)=-\sigma^{(1);xy}$, which can be expressed as
\begin{eqnarray}
  \sigma^{(1);xx} (\omega)  &=&
  \dfrac{1}{2}\sum_\tau \sigma^{(1);\tau\tau}(\omega)\,,\notag\\
  \sigma^{(1);yx}(\omega)&=&
  -\dfrac{i}{2}\sum_\tau \tau \sigma^{(1);\tau\tau}(\omega)\,.
 \label{eq:aalpha1}
\end{eqnarray}
The independent Cartesian components of the third order conductivity are
$\sigma^{(3);dxyy}$,
$\sigma^{(3);dyxy}$, and 
$\sigma^{(3);dyyx}$ with $d=x,y$. The other components can be obtained by
$\sigma^{(3);dxxx}=\sigma^{(3);dxyy}+\sigma^{(3);dyxy}+\sigma^{(3);dyyx}$
and by the symmetry $\{x\leftrightarrow y\}$. In this work, we
are interested in $\sigma^{(3);dxxx}$, which can be expressed as
\begin{eqnarray}
  \sigma^{(3);xxxx}  &=&  \frac{1}{4}\sum_{\tau}\left(\sigma^{(3);\tau\tau\tau\overline{\tau}}+
  \sigma^{(3);\tau\tau\overline{\tau}\tau}
  +
  \sigma^{(3);\tau\overline{\tau}\tau\tau}
  \right)\,,\notag\\
        {\sigma}^{(3);yxxx} 
        &=&-\frac{i}{4}\sum_{\tau}\tau\left(\sigma^{(3);\tau\tau\tau\overline{\tau}}+
        \sigma^{(3);\tau\tau\overline{\tau}\tau}
        +
        \sigma^{(3);\tau\overline{\tau}\tau\tau}
        \right)\,.\quad\quad
        \label{eq:aalpha2}
\end{eqnarray}
Obviously, the independent components $\sigma^{(1);xx}$,
$\sigma^{(3);xxyy}$, $\sigma^{(3);xyxy}$, and $\sigma^{(3);xyyx}$ are
nonzero regardless of the value of the magnetic
field, while those of 
$\sigma^{(1);xy}$, $\sigma^{(3);xyxx}$, $\sigma^{(3);xxyx}$, and
$\sigma^{(3);xxxy}$ are not zero only at nonzero
magnetic field. 

\subsection{Resonance and electron-hole symmetry}
\begin{figure}[tph]
  \centering
  \includegraphics[width=8cm]{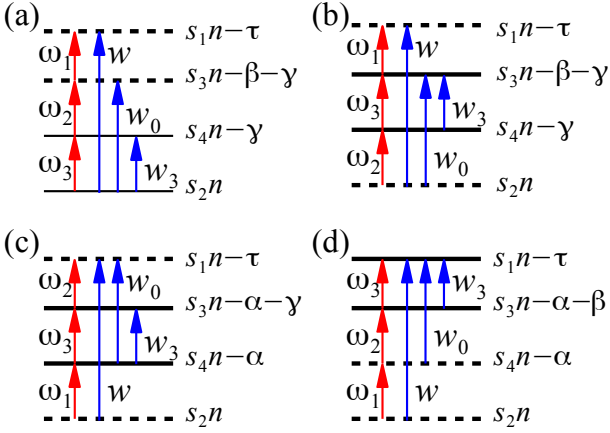}
  \caption{(color online) Optical transitions for  $\widetilde{\sigma}_{\nu}^{(3);\tau{\alpha}{\beta}{\gamma}}(w,w_0,w_3)$ with
    $w_3=\hbar\omega_3+i\Gamma$,
    $w_0=\hbar(\omega_2+\omega_3)+i\Gamma$,  and
    $w=\hbar(\omega_1+\omega_2+\omega_3)+i\Gamma$. The subfigures (a), (b), (c), and 
    (d) correspond to the four terms in Eq.~(\ref{eq:sigma3}),
    respectively. The horizontal lines indicate the
    LLs, where the indices ``$sn$'' are labeled at the right;
    the dashed lines mean the virtual states, while the two 
    solid lines stand for a pair of occupied and unoccupied states;
    the red arrows indicate the optical transitions with the involved
    photon frequency $\omega_i$ labeled aside; the blue arrows stand
    for the energies involved in the denominators.  }
\label{fig:illusnontrans}
\end{figure}

Here we discuss two general properties of the conductivity.
The first is related to the resonant transitions. 
At a finite magnetic field, all electronic levels are discrete, and
both the linear conductivity and the nonlinear conductivity posses
many resonant peaks. As an example, the transition diagrams for the
unsymmetrized third order 
conductivity $\widetilde{\sigma}^{(3);\tau\alpha\beta\gamma}(w,w_0,w_3)$ in
Eq.~(\ref{eq:sigma3}) are shown 
in Fig.~\ref{fig:illusnontrans}.
Because the indices $\tau\alpha\beta\gamma$ can only take the values
$\pm1$, each transition involves three photon energies, four bands
indices, and at most three and at least two Landau indices. Three
arrows associated with $w$, $w_0$, and $w_3$ correspond to the three
energy factors, in a form 
 \begin{equation}
   {\cal
   E}_{s_1s_2}(w,n,m)=w-(s_1\epsilon_{n+m}-s_2\epsilon_n)\,, \label{eq:Efactor}
 \end{equation}
 appearing
 in the denominator of the expression in Eq.~(\ref{eq:sigma3}). When
 ${\cal E}_{s_1s_2}(w,n,m)=0$, the optical transitions are in
 resonance, and third order nonlinear conductivity
 may diverge. The
 condition ${\cal E}_{ss}(w,n,m)=0$ determines the resonant intraband
 transition, while the condition ${\cal E}_{s\bar{s}}(w,n,m)=0$ determines the
 resonant interband transition. The expression of 
 $\widetilde{\sigma}^{(3);\tau\alpha\beta\gamma}(w,w_0,w_3)$ includes
 four full denominators, 
 \begin{eqnarray}
   &&{\cal E}_{s_1s_2}(w,n,\overline{\tau}){\cal
     E}_{s_3s_2}(w_0,n,\overline{\beta}+\overline{\gamma}) {\cal
     E}_{s_4s_2}(w_3,n,\overline{\gamma})\,,\notag \\
   &&{\cal E}_{s_1s_2}(w,n,\overline{\tau}){\cal
     E}_{s_3s_2}(w_0,n,\overline{\beta}+\overline{\gamma}) {\cal
     E}_{s_3s_4}(w_3,n-\beta,\overline{\gamma})\,,\notag \\
   &&{\cal E}_{s_1s_2}(w,n,\overline{\tau}){\cal
     E}_{s_1s_4}(w_0,n-\alpha,\overline{\beta}+\overline{\gamma}) {\cal
     E}_{s_3s_4}(w_3,n-\alpha,\overline{\gamma})\,,\notag \\
   &&{\cal E}_{s_1s_2}(w,n,\overline{\tau}){\cal
     E}_{s_1s_4}(w_0,n-\alpha,\overline{\beta}+\overline{\gamma}) {\cal
     E}_{s_2s_3}(w_3,n-\alpha-\beta,\overline{\gamma})\,.\notag
 \end{eqnarray}
Each of them is composed of three energy factors, depending on
field frequency $\omega_i$, magnetic field $B$, gap 
parameter $\Delta$, the Landau indices, and the polarization
of the incident light. By varying these parameters, one or more
${\cal E}$ can be zero and lead to a resonance. 
Usually each energy factor contributes one Lorentz-type
divergence as $(\delta E+i\Gamma)^{-1}$. Two energy factors in the 
 denominator may be the same, which could lead to a higher order divergence
as $(\delta E+i\Gamma)^{-2}$. Also, it can not be excluded that for
some special parameters ($n$, $s_i$, $\mu$, $\Delta$, $B$, and $\omega_i$)
all three energy factors simultaneously satisfy the resonant conditions, and lead
to a higher order divergence
\cite{Phys.Rev.Lett._108_255503_2012_Yao}.  We will use this analysis
for understanding the peaks in the spectra of the conductivities. 

The second is a chemical potential dependence of the conductivity that
is related to the electron-hole symmetry. 
For convenience, we explicitly show the chemical potential dependence as
$\sigma^{(1);\tau\alpha}_\nu(\omega,\mu)$,
$\sigma^{(3);\tau\alpha\beta\gamma}_\nu(\omega_1,\omega_2,\omega_3,\mu)$,
and $f_{\nu s n}(\mu)$, where the subscript $\nu$  indicates the
contribution from each valley. We use the relations in
Eqs.~(\ref{eq:xiconj}) and (\ref{eq:xivalley}), as well as
\begin{equation}
  f_{\nu s n}(\mu) = 1-f_{\overline{\nu}\bar{s} n}(-\mu)\,,
\end{equation}
where the energy relations between two valleys
$E_{\nu s n}=-E_{\overline{\nu}\overline{s}n}$ is used. Checking the 
expressions in Eqs.~(\ref{eq:sigma1}) and (\ref{eq:sigma3}) we can
directly obtain 
\begin{eqnarray}
  \sigma^{(1);\delta\alpha}_\tau(\omega,\mu) &=&
  \sigma_{\overline{\tau}}^{(1);\overline{\delta}\overline{\alpha}}(\omega,-\mu)\,,\\
  \sigma^{(3);\delta\alpha\beta\gamma}_\tau(\omega_1,\omega_2,\omega_3,\mu) &=&
  \sigma_{\overline{\tau}}^{(1);\overline{\delta}\overline{\alpha}\overline{\beta}\overline{\gamma}}(\omega_1,\omega_2,\omega_3,-\mu)\,.\quad
\end{eqnarray}
Using these relations and from Eqs.~(\ref{eq:aalpha1}) and
(\ref{eq:aalpha2}), the independent components
$\sigma^{(1);xx}$, $\sigma^{(3);xxyy}$, $\sigma^{(3);xyxy}$, and
$\sigma^{(3);xyyx}$ are even functions of the chemical potential,
while the other independent components $\sigma^{(1);xy}$,
$\sigma^{(3);xyxx}$, $\sigma^{(3);xxyx}$, and $\sigma^{(3);xxxy}$ are
odd functions of the chemical potential. Therefore, for intrinsic
graphene $\mu=0$ or GG, all transverse optical conductivities are
zero, {\it   e.g.}, $\sigma^{(1);xy}(\mu=0)=0$, for all temperature.  

In the following sections we discuss the linear
conductivity and nonlinear conductivities
associated with nonlinear phenomena including THG, NL, and FWM. For an
incident light $\bm E(t)=\bm E_\omega e^{-i\omega t} + c.c.$, the generated
nonlinear current for THG is
\begin{widetext}
\begin{eqnarray}
  \bm J_{\text{THG}}(t) &=& e^{-i3\omega t}\bm
  E_\omega\cdot\bm E_\omega  \left[
    \sigma^{(3);xxxx}(\omega,\omega,\omega) \bm E_\omega+
    \sigma^{(3);yxxx}(\omega,\omega,\omega)\hat{\bm z}\times \bm E_\omega \right]+ c.c.\notag\\
  &=& e^{-i3\omega t} E_\omega^+E_\omega^- \left[ 3\sigma^{(3);+++-}(\omega,\omega,\omega) E_\omega^+
    \hat{\bm e}^- + 3\sigma^{(3);---+}(\omega,\omega,\omega)E_\omega^-\hat{\bm e}^+\right]
 + c.c.\,;
\end{eqnarray}
for incident light $\bm E(t)=\bm E_{\omega_p}e^{-i\omega_pt} + \bm
E_{-\omega_s}e^{i\omega_s t} + c.c.$, the nonlinear current for FWM is
\begin{eqnarray}
  \bm J_{\text{FWM}}(t)  &=&
  e^{-(2\omega_p-\omega_s)t}\left\{\left[2\sigma^{(3);xxyy}(\omega_p,\omega_p,-\omega_s)\bm
    E_{\omega_p} + 2\sigma^{(3);xyxx}(\omega_p,\omega_p,-\omega_s)
    \bm E_{\omega_p}\times\hat{\bm z}\right]  \bm E_{\omega_p}\cdot\bm E_{-\omega_s} \notag\right.\\
  &&\left.+ \left[\sigma^{(3);xyyx}(\omega_p,\omega_p,-\omega_s) \bm
    E_{-\omega_s} + \sigma^{(3);xxxy}(\omega_p,\omega_p,-\omega_s)\bm
    E_{-\omega_s}\times \hat{\bm z}\right]\bm E_{\omega_p}\cdot\bm
  E_{\omega_p}\right\}+ c.c.\notag\\
 &=& e^{-(2\omega_p-\omega_s)t}\sum_\tau \hat{\bm e}^{\overline{\tau}}
  \left[\sigma^{(3);\tau\tau\tau\overline{\tau}}(\omega_p,\omega_p,-\omega_s)E_{\omega_p}^\tau     +
    2\sigma^{(3);\overline{\tau}{\tau}\overline{\tau}{\tau}}(\omega_p,\omega_p,-\omega_s)E_{\omega_p}^{\overline{\tau}}\right]E_{\omega_p}^\tau
  E_{-\omega_s}^{\overline{\tau}}+c.c.\,.  
\end{eqnarray}
\end{widetext}
By setting $\omega_s=\omega_p=\omega$, we can get the nonlinear
current for the NL process. Due to the
symmetry of graphene, there is no THG for single circularly polarized
incident light. We focus on the longitudinal current for incident
light along the $x$ direction, and note 
\begin{equation}
  \begin{array}{rcl}
  \sigma^{xxxx}_{\text{THG}}(\omega)&=&\sigma^{(3);xxxx}(\omega,\omega,\omega)\,,\\
  \sigma^{xxxx}_{\text{nl}}(\omega)&=&\sigma^{(3);xxxx}(\omega,\omega,-\omega)\,,\\
  \sigma_{\text{FWM}}^{xxxx}(\omega_p,\omega_s) &=&
  \sigma^{(3);xxxx}(\omega_p,\omega_p,\omega_s)\,.
  \end{array}
\end{equation}
The transverse conductivities are about two orders of magnitude
smaller than the longitudinal ones for most parameters, and are ignored
in this work.  

\section{Optical conductivities at weak magnetic
  field\label{sec:weakfield}}
We first consider the conductivities in a weak magnetic
field, where $\hbar\omega_c$ is much smaller than the relaxation
parameters, the thermal energy, and the involved photon energies. The
optical transitions between Landau levels can not be resolved, and it
is natural to treat the discrete LLs as continuous levels. 
As $B\to0$, the sum over the Landau level index $n$ in
Eqs.~(\ref{eq:sigma1}) and 
(\ref{eq:sigma3}) can be transformed into an integration over $x$ by the
substitution $\epsilon_n\to x$, $\epsilon_{n+\tau}\to x +
\tau(\hbar\omega_c)^2 F_0(x) $, and $w^{(\nu\tau)}_{s_1n+\tau_1;s_2
  n+\tau_2}\to F(x)+\tau_1(\hbar\omega_c)^2 F_1(x) +
\tau_2(\hbar\omega_c)^2 F_2(x)$, where the explicit expression
of $F_i(x)$ are not important for our next discussion. For linear conductivity
$\sigma^{(1);\tau\tau}(\omega)$, the contribution from the weak
magnetic field depends on $\tau(\hbar\omega_c)^2\propto \tau B$, 
thus we get
\begin{equation}
  \sigma^{(1);\tau\tau}(\omega) = S^{(1)}_0(\omega) -i S^{(1)}_1(\omega)  B \tau + S^{(1)}_2(\omega)  B^2 + \cdots\,.\label{eq:sigmaBexp}
\end{equation}
At weak magnetic field, the leading term of the linear conductivity is
\begin{eqnarray}
  \sigma^{(1);xx}(\omega) &\approx& S^{(1)}_0(\omega)\,,\\
  \sigma^{(1);xy}(\omega) &\approx& S^{(1)}_1(\omega) B\,.
\end{eqnarray}
Similarly, the third order conductivities can be approximated as
\begin{eqnarray}
  \sigma^{(3);xxxx}(\omega_1,\omega_2,\omega_3) &\approx& S^{(3)}_0(\omega_1,\omega_2,\omega_3)\,,\\
  \sigma^{(3);yxxx}(\omega_1,\omega_2,\omega_3) &\approx& S^{(3)}_1(\omega_1,\omega_2,\omega_3) B\,.
\end{eqnarray}

The study of this limit has two aims. One is
a validity check by comparing our conductivities of doped graphene at very weak magnetic field with
those at zero magnetic field, which are already presented in the
literature. The other is to approach the third 
order conductivities for gapped graphene at zero magnetic field.

\subsection{Comparison with literature}
In the absence of the magnetic field, the optical conductivities of DG
have been  systematically studied both analytically
\cite{NewJ.Phys._16_53014_2014_Cheng,Phys.Rev.B_91_235320_2015_Cheng,Phys.Rev.B_93_85403_2016_Mikhailov}
and numerically \cite{Phys.Rev.B_92_235307_2015_Cheng}. In the linear
dispersion approximation and describing the relaxation in a
phenomenological way, analytic expressions are found for linear
\cite{J.Appl.Phys._103_64302_2008_Hanson}, second order
\cite{Phys.Rev.B_94_195442_2016_Wang,Sci.Rep._7_43843_2017_Cheng,Phys.Rev.B_95_35416_2017_Rostami}, and third order conductivities
\cite{Phys.Rev.B_91_235320_2015_Cheng,Phys.Rev.B_93_85403_2016_Mikhailov}. For GG, the linear conductivity has analytic expression for transitions
around the Dirac points, and some of the nonlinear conductivities have
been numerically extracted  \cite{Phys.Rev.B_92_235307_2015_Cheng}. All
those conductivities are based on a plane wave basis. In this work, these conductivities are
considered from a different approach, based on the LLs, and it is
interesting to compare them.  

\begin{widetext}
\begin{figure*}[ht]
  \centering
  \includegraphics[width=5.5cm]{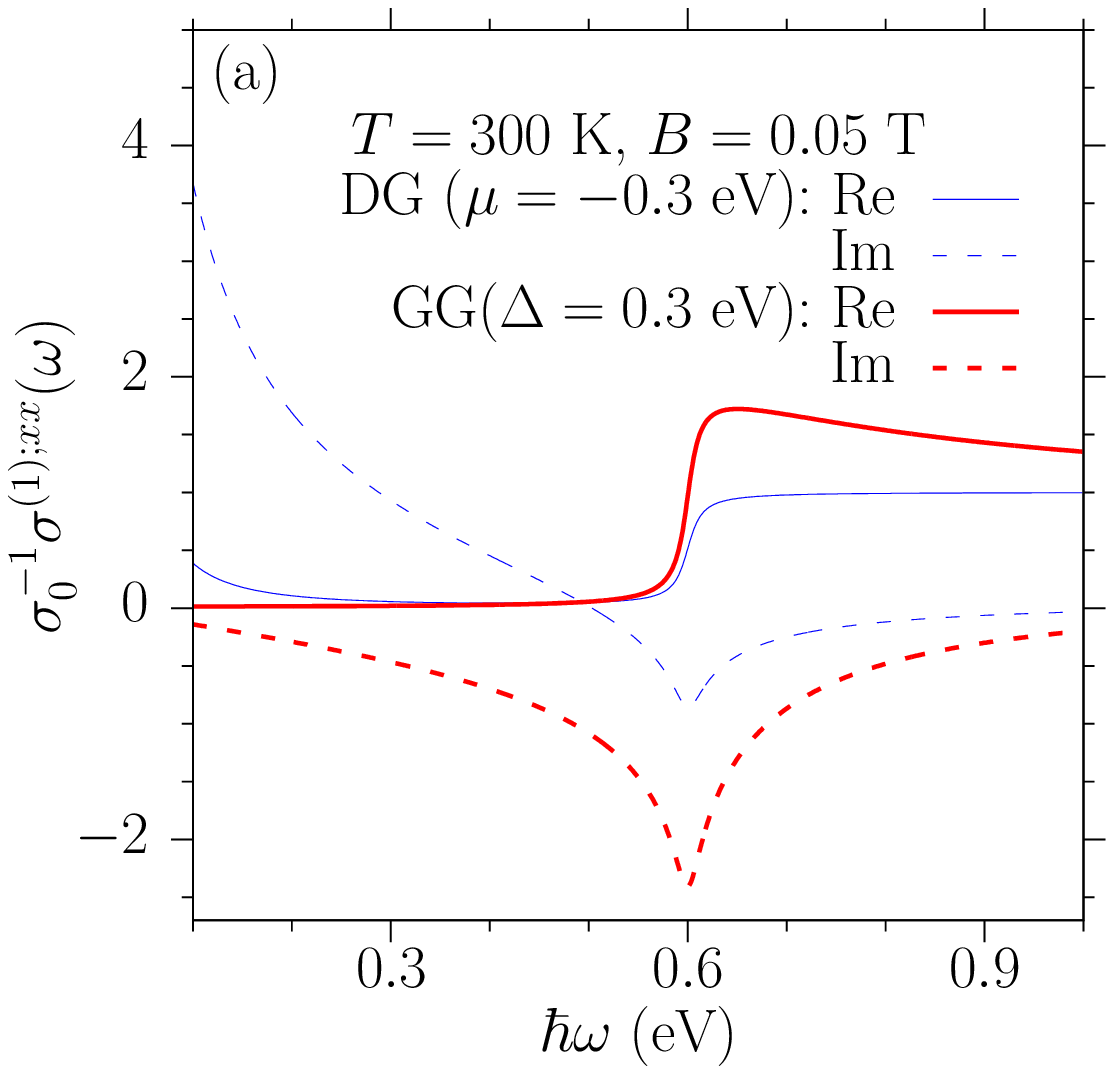}
  \includegraphics[width=5.5cm]{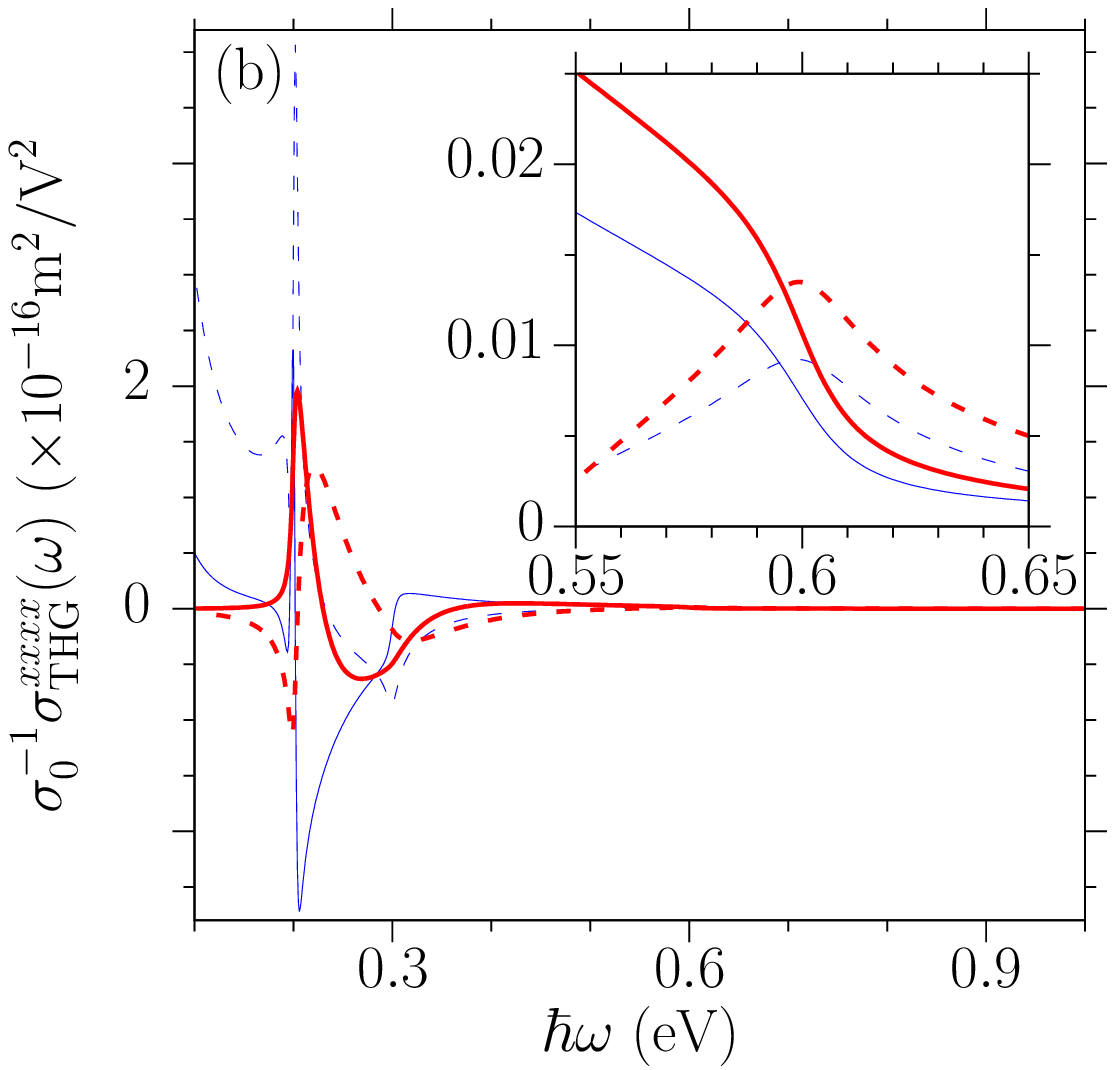}\\  
  \includegraphics[width=5.5cm]{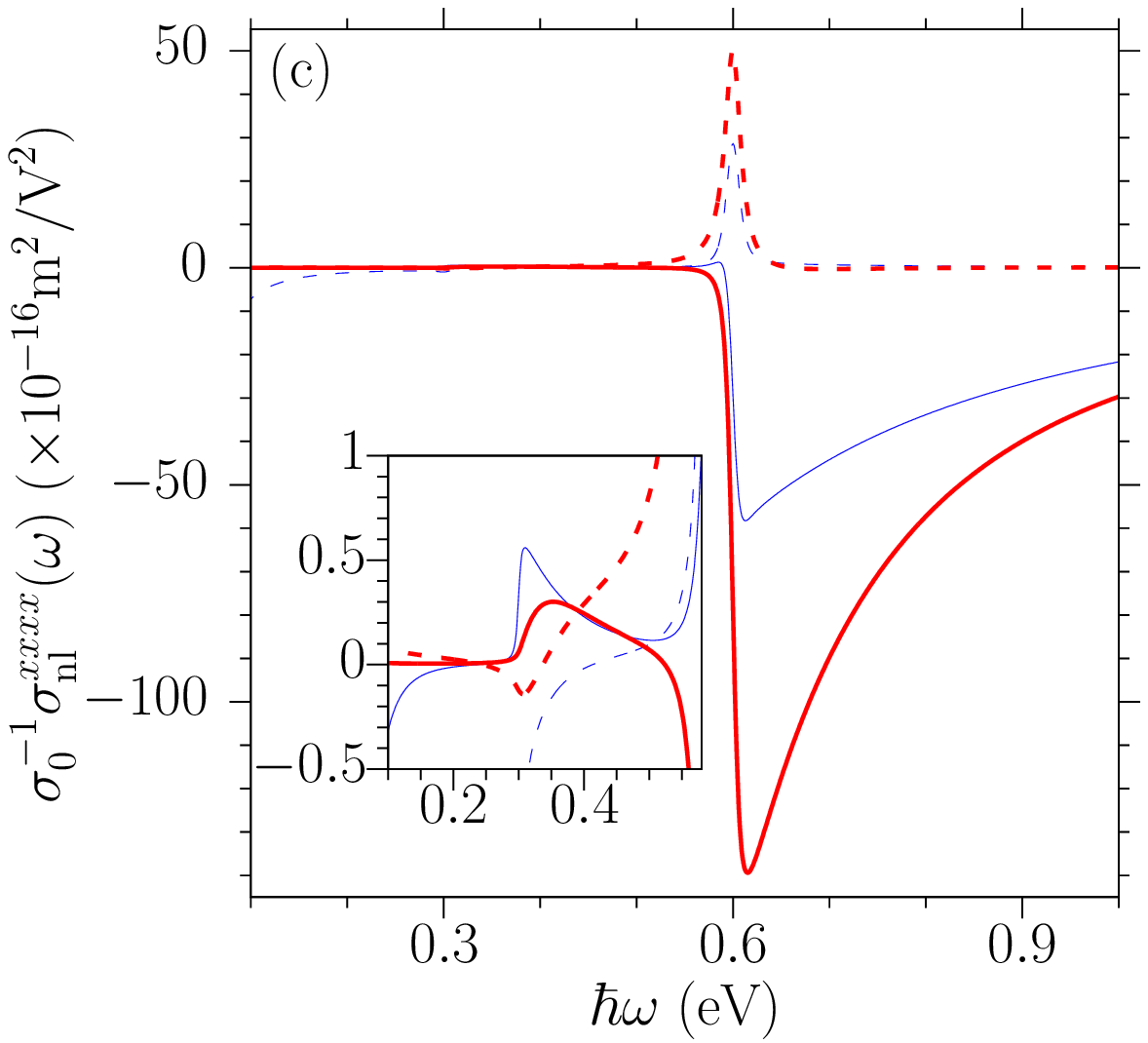}
  \includegraphics[width=5.5cm]{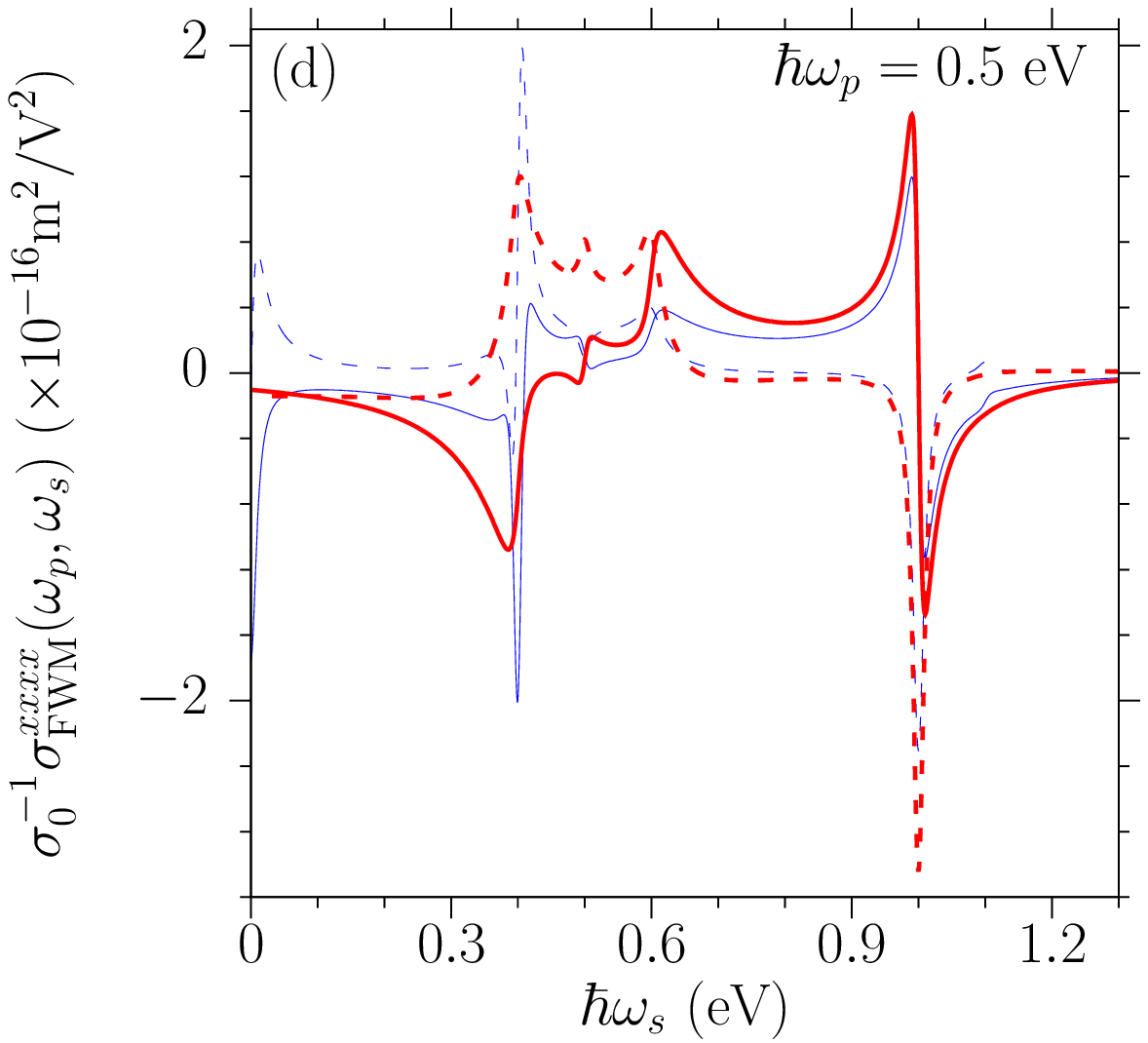}
  \caption{(color online) Spectra of optical conductivities for DG at
    $\mu=-0.3$~eV (blue thin curves) and GG at $\Delta=0.3$~eV
    (red thick curves). (a) $\sigma^{(1);xx}(\omega)$,    (b)
    $\sigma^{xxxx}_{\text{THG}}(\omega)$, (c) $\sigma^{xxxx}_{\text{nl}}(\omega)$, (d)
    $\sigma^{xxxx}_{\text{FWM}}(\omega_p,\omega)$. The solid (dashed)
    curves are for the real
    (imaginary) parts, respectively. The insets in (b) and (c) show
    the details in given range of the photon energies. Other parameters are $B=0.05$~T, $T=10$~K, and
    $\Gamma=10$~meV.} 
  \label{fig:spectraB0}
\end{figure*}
\end{widetext}

Our calculations are done for parameters $B=0.05$~T. Here the
treatment of $0.05$~T as weak field can be clearly seen in
section~\ref{sec:strongfield}.   Other parameters are
$\Gamma=10$~meV, the temperature $T=10$~K, and the Landau index $n$
is taken as $n<N_c$ with a cutoff $N_c$. For high photon energy, large
$N_c$ is required for the convergence of the conductivities. Because
the photon energy can be as high as around 1~eV, we 
choose $N_c=2\times 10^6$ (corresponding to an energy cutoff
$\sim 11$~eV) for the calculation of linear conductivity, and
$N_c=10^6$ (the energy cutoff is $\sim8$~eV) for the nonlinear
conductivities.  The spectra of the conductivities of DG are plotted in
Fig.~\ref{fig:spectraB0} as thin curves for $\sigma^{(1);xx}$,
$\sigma^{xxxx}_{\text{THG}}$, 
$\sigma^{xxxx}_{\text{nl}}$, and $\sigma^{xxxx}_{\text{FWM}}$; and the
linear conductivity of GG $\sigma^{(1);xx}$ is shown in
Fig.~\ref{fig:spectraB0} (a) as~thick curves.
The obtained curves overlap with those from analytic
expressions~\cite{Phys.Rev.B_91_235320_2015_Cheng,Phys.Rev.B_92_235307_2015_Cheng},
confirming the equivalence between the LL basis and the plane wave
basis.

The comparison can be further extended to the linear
conductivity  $\sigma^{(1);xy}(\omega)$ or $S_1^{(1)}(\omega)$, which
is related to the second order conductivities induced by magnetic
dipole interactions. In DG, the second order
current
\cite{Sci.Rep._7_43843_2017_Cheng} can be expressed as
\begin{eqnarray}
  \bm J^{(2)}(\bm r,t) &=& 2 \int\frac{d\bm q_1 d\bm q d\omega_1
    d\omega}{(2\pi)^6} e^{-i(\omega_1+\omega)t+i(\bm q_1+\bm
    q)\cdot\bm r}\notag\\
  &&\times \Big\{ S_M^{xxyy}(\omega_1,\omega) [\bm q_1\times
    \bm E(\bm q_1\omega_1)]\times \bm E(\bm q\omega)\notag\\
  && +S_Q^{xyxy}(\omega_1,\omega) \bm
  q_1\cdot\bm E(\bm q_1\omega_1)\bm E(\bm q\omega)\notag\\
  &&+
  S_Q^{xxyy}(\omega_1,\omega) [\bm E(\bm q_1\omega_1) \bm
    q_1\cdot\bm E(\bm q\omega) \notag\\
    && + \bm q_1\bm E(\bm q_1\omega_1) 
    \cdot\bm E(\bm q\omega)]\Big\}\,.
\end{eqnarray}
In the linear dispersion approximation, the coefficients $S^{xxyy}_M$,
$S^{xyxy}_Q$, and $S^{xxyy}_Q$ have analytic expressions
\cite{Sci.Rep._7_43843_2017_Cheng}, where the 
term involving $S_M^{xxyy}$ gives the contribution induced by the magnetic
dipole interaction. Using the relation $\bm B(\bm q_1\omega_1) =
\frac{1}{\omega_1}\bm q_1\times \bm 
E(\bm q_1\omega_1)$ and taking the limit $\bm q_1,\omega_1,\bm q\to0$, we
can recover the uniform magnetic field,  and obtain
\begin{eqnarray}
  S^{(1)}_1(\omega) &=&
  \lim\limits_{\omega_1\to0}\left[2\omega_1S_M^{xxyy}(\omega_1,\omega)
    - 2\omega_1S_M^{xxyy}(-\omega_1,\omega)\right]\notag\\
  &=& -\text{sgn}(\mu) \sigma_0 \frac{2e(\hbar v_F)^2}{\pi\hbar}
  \frac{1}{(\hbar\omega)^2}\left(\frac{8\mu^2}{w^2-4\mu^2}\right.\notag\\
    && \left.+ \frac{8\mu^2}{\Gamma^2+4\mu^2}\frac{2w-i\Gamma}{w}\right)\,,
\end{eqnarray}
with $w=\hbar\omega+i\Gamma$ and the universal conductivity $\sigma_0=e^2/(4\hbar)$. In the absence of the relaxation,
$S_1^{(1)}(\omega)$ agrees with the calculated
$\sigma^{(1);xy}(\omega)$ very well. 

These two approaches show excellent agreement, which provides a good
validity check of the approach based on the LLs. Furthermore, the nonlinear
conductivities of GG in the absence of the magnetic field, which has not
been systematically studied before, can be obtained from
Eqs.~(\ref{eq:sigma1}) and (\ref{eq:sigma3})  at very weak magnetic
field $B=0.05$~T.  

\subsection{Third order conductivities of GG}
We calculate the optical conductivity of GG for a gap
parameter $\Delta=0.3$~eV and a chemical potential $\mu=0$~eV. Other
parameters are  $B=0.05$~T, $T=10$~K, and 
$\Gamma=10$~meV, which are the same as those in previous section. The
largest energy difference between adjacent Landau levels is
$\epsilon_1-\epsilon_0\sim0.1$~meV, which is much smaller than the
relaxation parameter $\Gamma$, and the discreteness of the levels can be smeared out by the
relaxation parameters. In Fig.~\ref{fig:spectraB0} we show the
calculated conductivities of GG in thick curves. Because the system
has a nonzero gap $2\Delta=0.6$~eV and it is not doped, both the
linear and nonlinear conductivities are insensitive to
temperature.  
In DG, the chemical potential $\mu$ can be used to tune the 
optical nonlinearity because it acts like a chemical potential induced
gap for 
interband transitions, and leads to resonant  responses. Analogically,
the real gap $2\Delta$ in GG is expected to play similar role. Here we
discuss their similarities and differences.

For simplicity we first look at the linear conductivity. As we
discussed in the previous section, our calculations are in agreement
with the analytic expressions \cite{Phys.Rev.B_92_235307_2015_Cheng}.
Our numerical results are shown in Fig.~\ref{fig:spectraB0}(a). With
photon energy increasing from 
0.1~eV to 1~eV, the real part of the linear conductivity is zero for
about $\hbar\omega<2\Delta$, quickly turns on for 
$\hbar\omega\sim 2\Delta$, and then decreases with photon energy to
$\sigma_0=e^2/(4\hbar)$. The imaginary part is negative 
for all photon energies, which shows the dielectric properties of a
gapped graphene. Its magnitude increases from zero at
$\hbar\omega=0$ (from the analytic expression) to a maximum value around
$\hbar\omega\sim2\Delta$, and then decreases.
Compared to the conductivity of DG at $\mu=\Delta$, there are two main
differences: (i) The Drude conductivity disappears because the thermal
excited carriers can be ignored, and thus the gapped graphene acts as a 
dielectric material. (ii) The value of the real part at the photon
energy around the gap is larger than that of DG, which can be
attributed to the larger values of the dipole matrix elements shown in
Fig.~\ref{fig:xienvp}; similarly, the peak 
value of its imaginary part is also larger than that of  DG.

Now we turn to the nonlinear conductivities of GG. The spectra of
$\sigma^{xxxx}_{\text{THG}}(\omega)$ for THG,
$\sigma^{xxxx}_{\text{nl}}(\omega)$ for NL, and
$\sigma^{xxxx}_{\text{FWM}}(\omega_p,\omega_s)$ for FWM are plotted in
Fig.~\ref{fig:spectraB0} (b), (c), and (d), respectively. All 
spectra show complicated dependence on the photon energy, and they include many
resonant peaks. For
$\sigma^{xxxx}_{\text{THG}}(\omega)$  in Fig.~\ref{fig:spectraB0} (b),
its value is almost zero at
 $\hbar\omega=0.1$~eV. With increasing the photon energy, its
magnitude increases and reaches a resonant value around $2\times
10^{-16}\times \sigma_0$ m$^2$/V$^2$ at photon energy
$\hbar\omega=0.2$~eV. Then 
its magnitude generally decreases with the photon energy, and shows
two more resonant features around $\hbar\omega=0.3$~eV and $0.6$~eV,
but with smaller peak values. The fine
structures of $\sigma^{xxxx}_{\text{THG}}$ are located at $\hbar\omega\sim
0.2$~eV, $0.3$~eV, and $0.6$~eV. For
$\sigma^{xxxx}_{\text{nl}}(\omega)$  in Fig.~\ref{fig:spectraB0} (c),
we first look at the real part
$\text{Re}[\sigma^{xxxx}_{\text{nl}}(\omega)]$. Its value remains
close to zero for photon energies $\hbar\omega<0.3$~eV, increases
quickly to reach a peak value 
$\sim0.3\times 10^{-16}\times \sigma_0$ m$^2$/V$^2$ around
$\hbar\omega\sim0.35$~eV, and then decreases to cross zero around
$\hbar\omega\sim0.53$~eV. For greater photon energies,
the real part reaches a valley with a negative value $\sim -140\times
10^{-16}\times \sigma_0$ m$^2$/V$^2$ at $\hbar\omega\sim 0.61$~eV. At
higher photon energies, its value increases monotonically towards
zero. The imaginary part
$\text{Im}[\sigma^{xxxx}_{\text{nl}}(\omega)]$ decreases from $\sim0.07\times 10^{-16}\times \sigma_0$ m$^2$/V$^2$ at
$\hbar\omega=0.1$~eV to a value $\sim -0.15\times
10^{-16}\times \sigma_0$ m$^2$/V$^2$ at $\hbar\omega\sim 0.31$~eV,
then it increases to a peak with value $\sim 50\times
10^{-16}\times \sigma_0$ m$^2$/V$^2$ at $\hbar\omega\sim 0.6$~eV, and
finally decreases to small values for higher photon energy. The fine
structures are located at $\hbar\omega\sim 0.3$~eV and $0.6$~eV.  The
spectrum of $\sigma^{xxxx}_{\text{FWM}}(\omega_p,\omega_s)$  in
Fig.~\ref{fig:spectraB0} (d) shows even more complicated structures,
which are separated by photon energies at $\hbar\omega\sim0.4$~eV,
$0.5$~eV, $0.6$~eV, and $1.0$~eV. The values are also at the order of
magnitude of $10^{-16}\times \sigma_0$ m$^2$/V$^2$.

All these features can also be found in the spectra of conductivities
of DG (thin curves in the same figure), and are all induced by the
resonant interband transitions. Some of them are related to the
energy of the gap ($E_g=2\Delta$ for GG or $E_g=2|\mu|$ for DG), with
matching the involved photon energies or their sum with the gap energy. For
THG, the conditions are 
$E_g=\hbar\omega$,  $2\hbar\omega$, or $3\hbar\omega$; for NL, they are
$E_g=\hbar\omega$ or $E_g=2\hbar\omega$; and for FWM, they are
$E_g=2\hbar\omega_p-\hbar\omega_s$ and $\hbar\omega_s$. The other two
resonant features in FWM occur at the conditions 
$\hbar\omega_s=\hbar\omega_p$  and
$\hbar\omega_s=2\hbar\omega_p$. They arise from the optically excited free
carriers.  In fact, for $\hbar\omega_s=\hbar\omega_p$, FWM reduces to NL
$\sigma^{xxxx}_{\text{nl}}(\omega_p)=\sigma^{xxxx}_{\text{FWM}}(\omega_p,\omega_p)$. In
our calculation, the NL is finite and the corresponding structure in
the spectra of FWM does not change too much. The other condition
$\hbar\omega_s=2\hbar\omega_p$ corresponds to two-color coherent
current injection, which show a Lorentz-type divergence as
$(\hbar \delta\omega+i\Gamma)^{-1}$ for small
$\hbar\delta\omega=2\hbar\omega_p-\hbar\omega_s$. We conclude that these fine
structures of the spectra in GG also come from the interband resonant
transitions, and  the chemical potential in DG
and the gap parameter in GG do have similar role for the interband
transition. The values of the resonant peaks of DG and GG are of 
the same order of magnitude, but 
differ by a factor of two or three. Some of them are larger and
sharper  for DG than
those for GG,  such as  the resonance at $0.2$~eV for THG, that at $0.3$~eV for NL,
and that at $0.4$~eV for FWM. In general, other peaks for DG are lower
than those for GG, similar to that of the linear conductivity. 

The differences between the conductivities of DG and GG are also
obvious, especially around the zero photon energies. All conductivities of DG show a Drude-like contribution, which tends to
diverge for zero photon energy (or large finite value with the
inclusion of relaxation). In the relaxation free case, the third order
nonlinear conductivities of DG behave as $\sigma^{xxxx}\propto
\omega^{-4}$. However, all conductivities of GG at low
photon energies give small values; this is consistent with a general
conclusion for cold and clean dielectric materials. Aversa and Sipe
\cite{Phys.Rev.B_52_14636_1995_Aversa} have 
theoretically shown that for a cold and clean 
dielectric material, the value of $\chi^{(3)}$ should be zero as all
frequencies go to zero, which
means $\sigma^{xxxx}_{\text{THG}}(0)=0$ and
$\sigma^{xxxx}_{\text{nl}}(0)=0$. 

\begin{widetext}
\begin{figure*}[htb]
  \centering
  \includegraphics[height=5.cm]{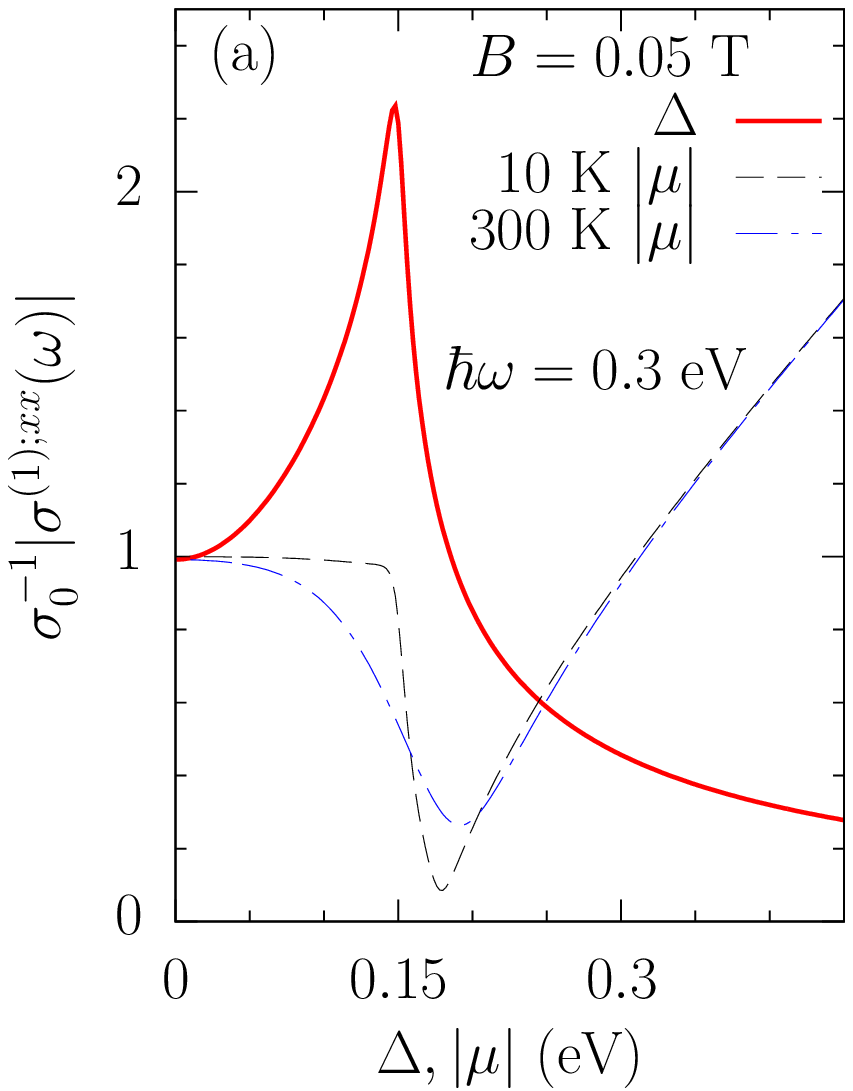}
  \includegraphics[height=5.cm]{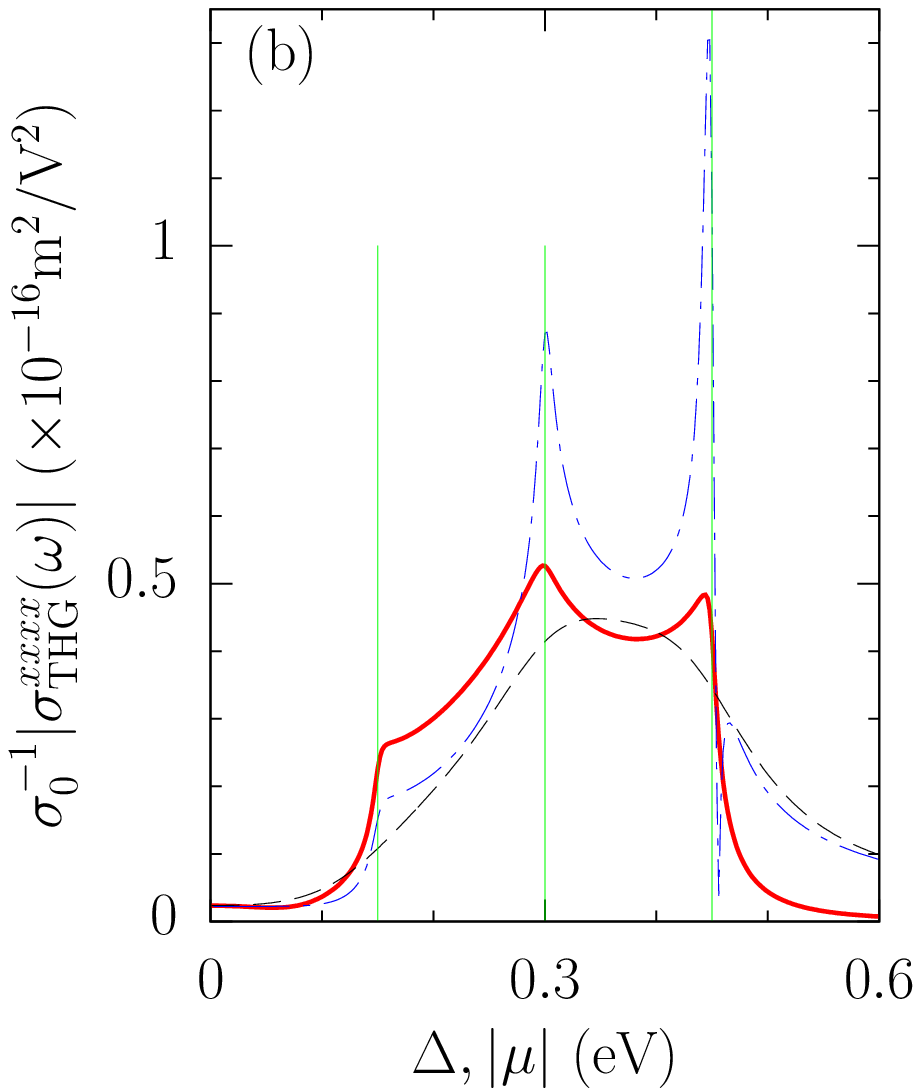}
  \includegraphics[height=5.cm]{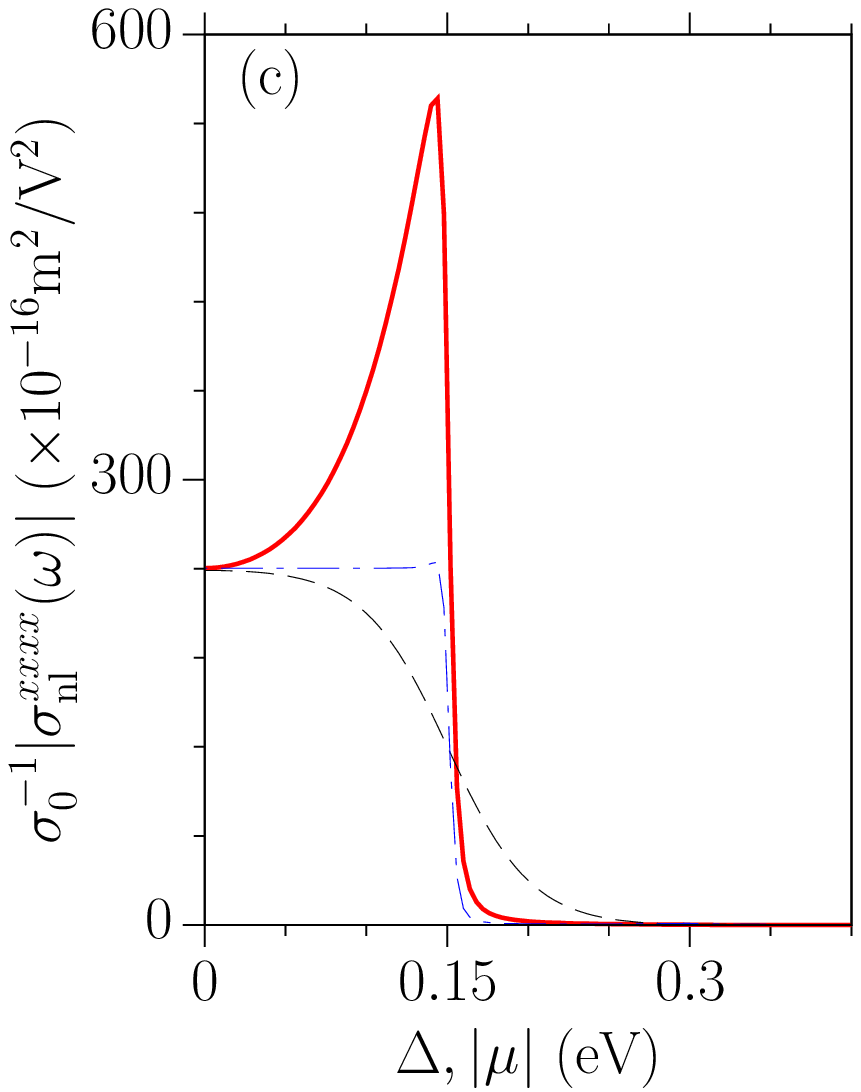}
  \includegraphics[height=5cm]{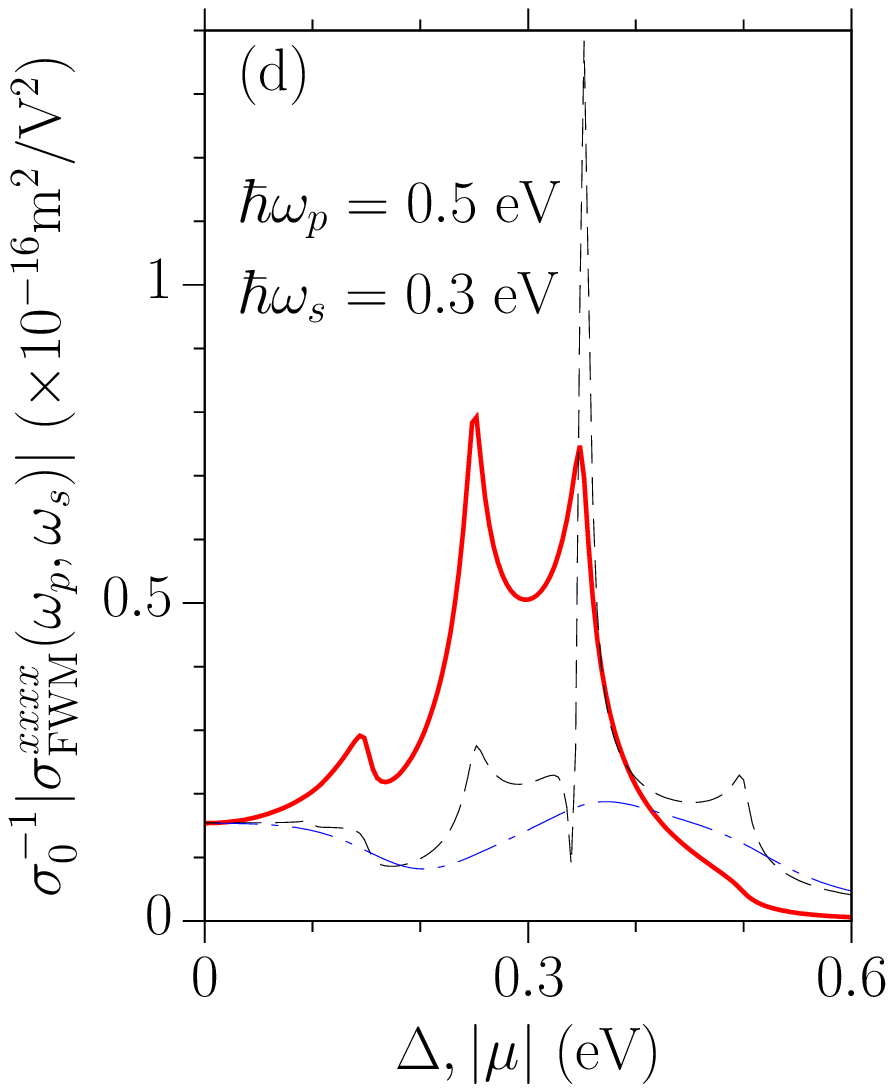}
  \caption{(color online) The $|\mu|$ ($\Delta$) dependence of 
    optical conductivities of DG (GG). (a) $|\sigma^{(1);xx}(\omega)|$, (b) $|\sigma^{xxxx}_{\text{THG}}(\omega)|$, (c)
    $|\sigma^{xxxx}_{\text{nl}}(\omega)|$, (d) 
    $|\sigma^{xxxx}_{\text{FWM}}(\omega_p,\omega_s)|$. The photon
    energies are chosen as $\hbar\omega=\hbar\omega_s=0.3$~eV,
    $\hbar\omega_p=0.5$~eV. Other parameters are $B=0.05$~T and $\Gamma=10$~meV. }
  \label{fig:spectraB}
\end{figure*}
\end{widetext}

This distinction can be further understood from the peaks in
Fig.~\ref{fig:spectraB}, where the absolute 
values of conductivities are plotted with varying $\mu$ for DG
or $\Delta$ for GG.~The
 photon energies are chosen as
$\hbar\omega=\hbar\omega_s=0.3$~eV and $\hbar\omega_p=0.5$~eV, the
magnetic field is $B=0.05$~T, and the temperature is $T=10$~K as well as
$300$~K. Note that for a finite gap $\Delta\gg k_BT$, all
conductivities of GG are insensitive to the temperature; 
however, temperature remarkably affects those of DG, even
smearing out some peaks, as shown in Fig.~\ref{fig:spectraB} for
$T=10$~K and $300$~K. 

\section{Magnetic field dependence of optical
  conductivities\label{sec:strongfield}} 
In this section we examine how the magnetic field affects
optical conductivities $\sigma^{(1);xx}(\omega)$,
$\sigma^{xxxx}_{\text{nl}}(\omega)$, 
$\sigma^{xxxx}_{\text{THG}}(\omega)$, and
$\sigma^{xxxx}_{\text{FWM}}(\omega_p,\omega_s)$. The calculations are
performed for different 
gap parameters for GG and chemical potential for DG at a relaxation
parameter $\Gamma=10$~meV, temperature $T=10$~K, and fixed photon
energies $\hbar\omega=\hbar\omega_s=0.3$~eV and
$\hbar\omega_p=0.5$~eV.  The cutoff of the Landau index $N_c$ is
chosen to satisfy the cutoff energy to be at the order of 10~eV.

\subsection{Results of GG}
\begin{widetext}
\begin{figure*}[ht]
  \centering
  \includegraphics[width=6cm]{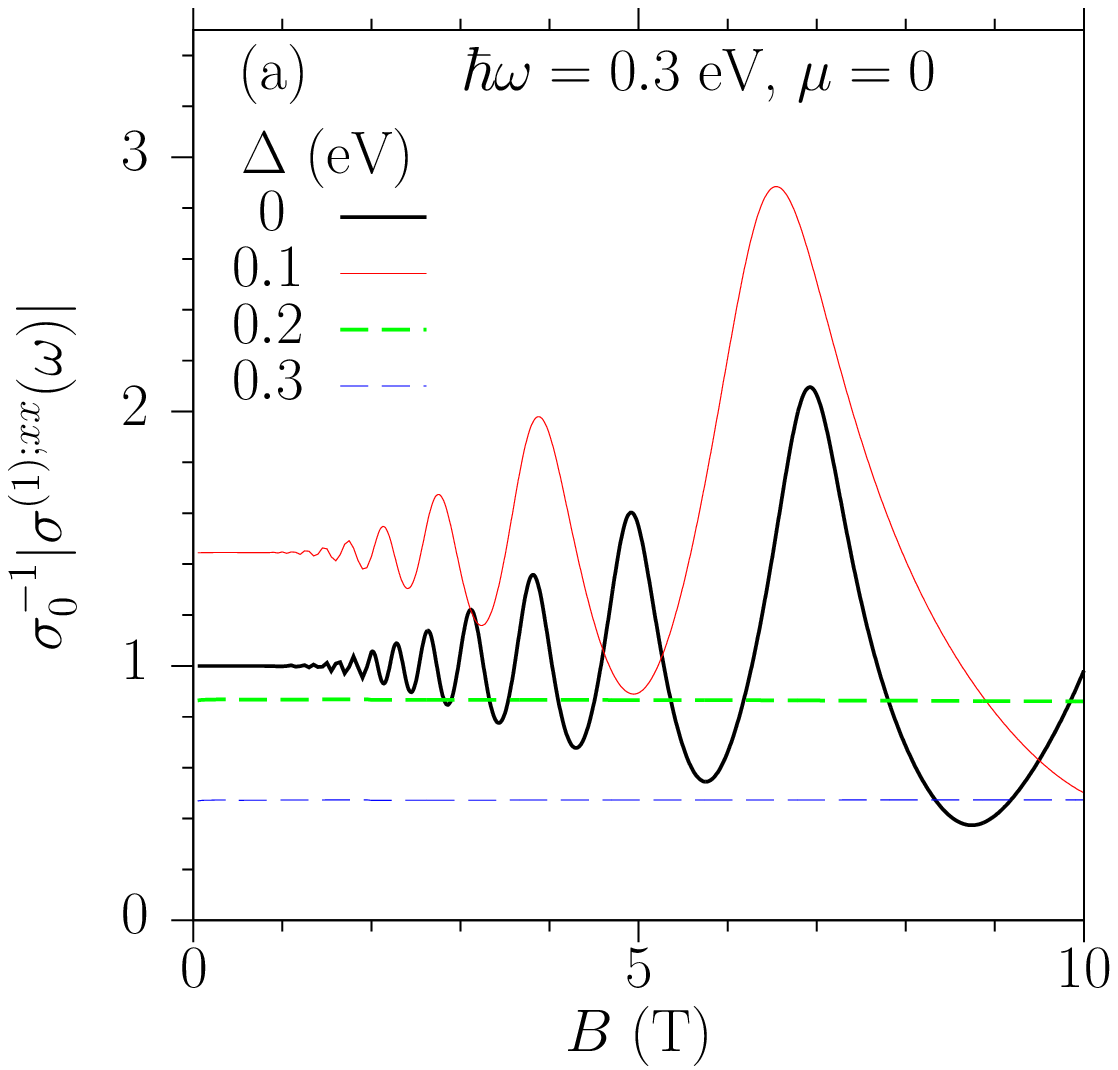}~~
  \includegraphics[width=6cm]{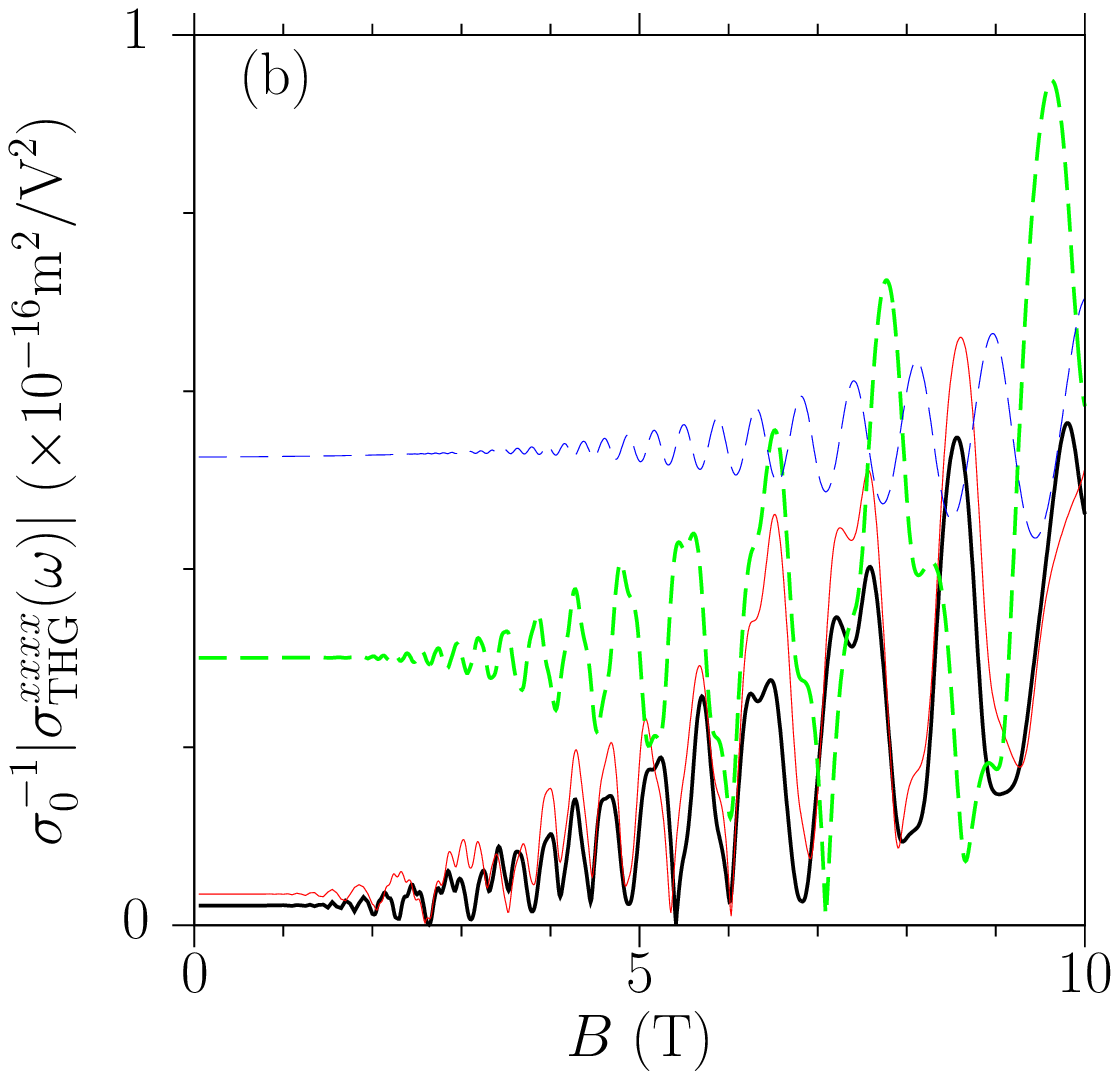}\\
  \includegraphics[width=6cm]{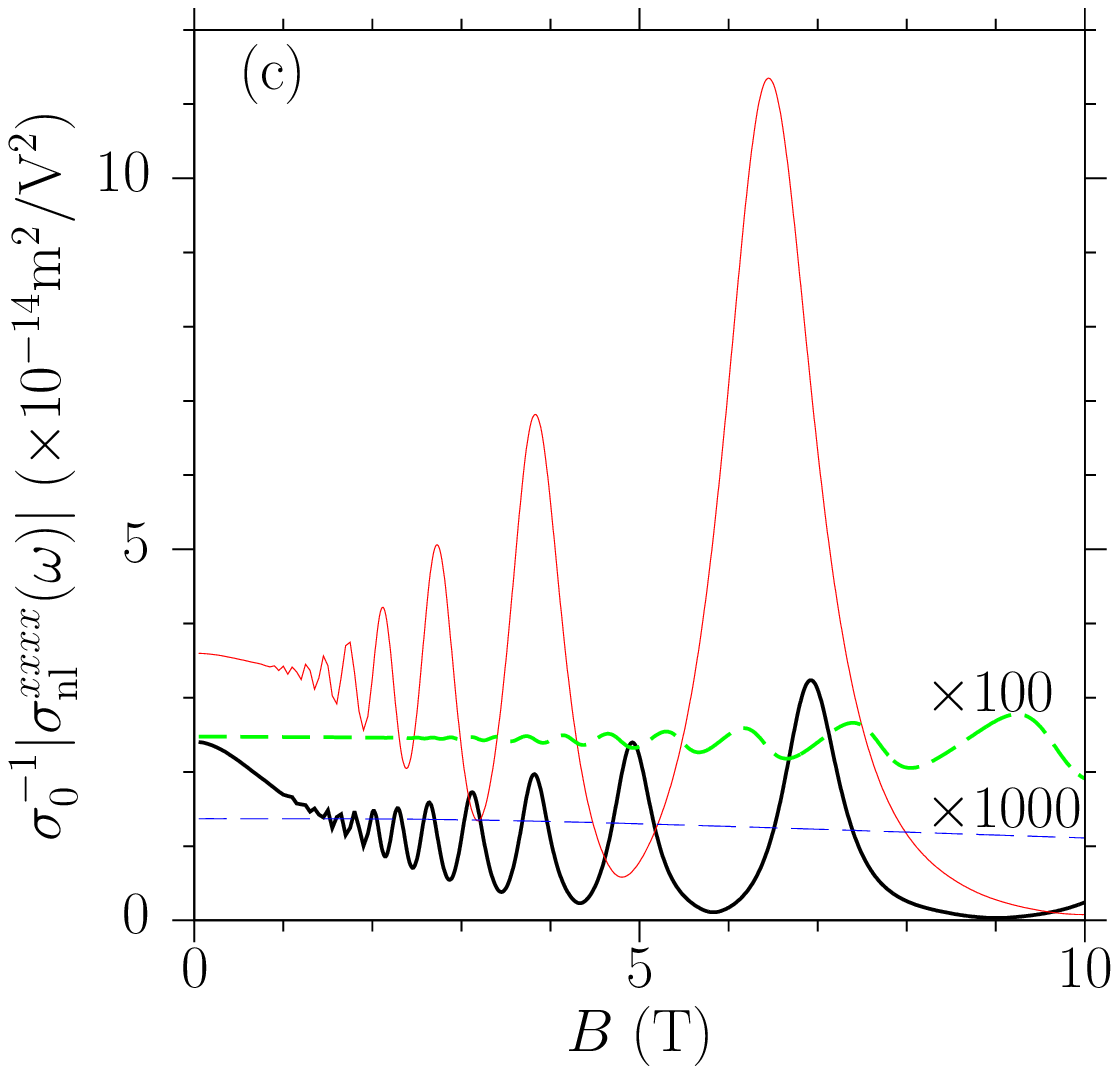}~~
  \includegraphics[width=6cm]{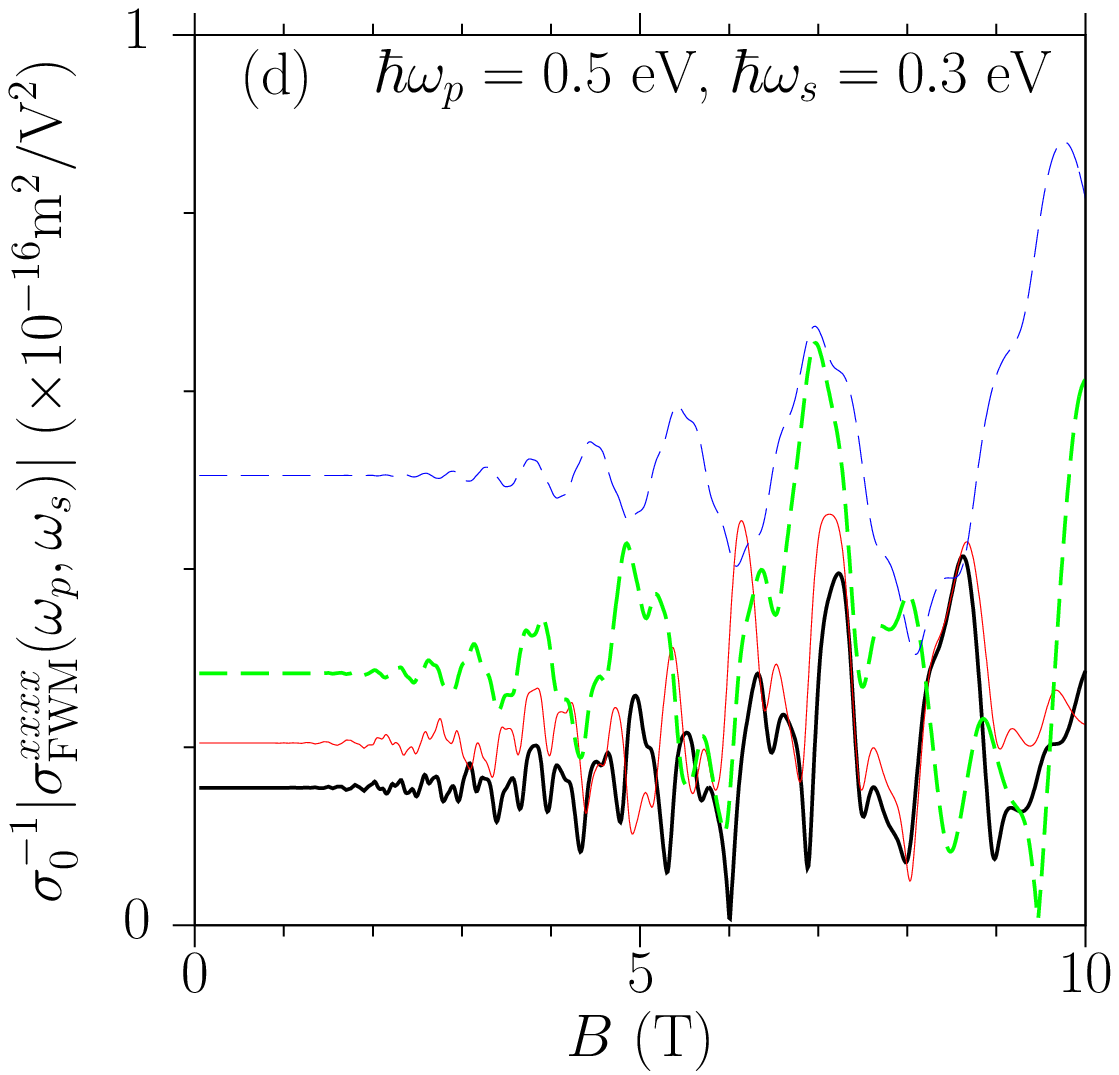}
  \caption{(color online) Magnetic field dependence of the absolute values of optical
    conductivities of 
    GG. (a) $|\sigma^{(1);xx})\omega)|$, (b)
    $|\sigma^{xxxx}_{\text{THG}}(\omega)|$,
    (c) $|\sigma^{xxxx}_{\text{nl}}(\omega)|$,
    (d) $|\sigma^{xxxx}_{\text{FWM}}(\omega_p,\omega_s)|$. The
    parameters are $\hbar\omega=\hbar\omega_s=0.3$~eV,
    $\hbar\omega_p=0.5$~eV, $\Gamma=10$~meV, $\mu=0$, and $T=10$~K. In
    (b), the curves for $\Delta=0.2$ and $0.3$~eV are scaled by 100
    and 1000 times, respectively.}
  \label{fig:gapCB}
\end{figure*}
\end{widetext}

In  Fig.~\ref{fig:gapCB}, we present the absolute values of
conductivities $\sigma^{(1);xx}(\omega)$, $\sigma^{xxxx}_{\text{nl}}(\omega)$,
$\sigma^{xxxx}_{\text{THG}}(\omega)$, and
$\sigma^{xxxx}_{\text{FWM}}(\omega_p,\omega_s)$ of 
a gapped graphene with varying the magnetic field $B\in[0.05,10]$~T for
gap parameters $\Delta=0$, $0.1$, $0.2$, and $0.3$~eV. We note that the
transverse components of $\sigma^{(1);xy}$, $\sigma^{(3);xyxx}$,
$\sigma^{(3);xxyx}$, and $\sigma^{(3);xxxy}$ are all zero for GG ($\mu=0$). 
 
At a magnetic field $B=0.05$~T, all conductivities have been
discussed in the previous section, and the gap dependence is shown
in Fig.~\ref{fig:spectraB}. In Figs.~\ref{fig:gapCB} (a, b, d),
$\sigma^{(1);xx}(\omega)$, $\sigma^{xxxx}_{\text{THG}}(\omega)$, and 
$\sigma^{xxxx}_{\text{FWM}}(\omega_p,\omega_s)$ show very weak
dependence on the magnetic field for $B<B_c$, where $B_c$
depends on the optical conductivity and 
the gap parameter. The value of $B_c$ increases with the increase of
the gap parameter. For $\sigma^{(1);xx}(\omega)$, $B_c$ is
about $2$~T for both $\Delta=0$ and $0.1$~eV, while it is not less
than $10$~T for $\Delta=0.2$ and $0.3$~eV. For
$\sigma^{xxxx}_{\text{THG}}(\omega)$ and
$\sigma^{xxxx}_{\text{FWM}}(\omega_p,\omega_s)$, $B_c$ is about $1$, 
$1.5$, $2$, and $3$~T for $\Delta=0$, $0.1$, $0.2$, and $0.3$~eV,
respectively. Obviously, for these three conductivities, the
coefficients of $B^2$ term in Eq.~(\ref{eq:sigmaBexp}) are negligible.
However, $\sigma^{xxxx}_{\text{nl}}$ shows different behavior, where
its absolute values for  $\Delta=0$ and $0.1$~eV decrease obviously
with increasing the magnetic field. This implies the contribution from
the $B^2$ term has to be taken into account. However, for the other two
$\Delta=0.2$ and $0.3$~eV, such dependence is very weak. By checking 
the optical transitions in Eq~(\ref{eq:sigma3}), $B^2$ term may be important in the 
interference process between two resonant transitions induced by
$\omega$  and $\omega+\omega-\omega$ (see Fig.2 (b) in
Ref.[\onlinecite{NewJ.Phys._16_53014_2014_Cheng}]), which exist only 
for $\Delta\le 0.15$~eV.

For $B>B_c$, the conductivities oscillate with the magnetic
field, and show the following features: (1) For each
 conductivity at a given $\Delta$, the oscillations are not a
periodic function of magnetic field; instead, the change in the field between neighboring oscillation peaks (hereafter noted as an oscillation
period) increases with the magnetic field. (2) When varying 
$\Delta$, both the period and the peak position of 
the oscillations change, especially for the nonlinear conductivities.
The oscillations of $\sigma^{xxxx}_{\text{THG}}$ and
$\sigma^{xxxx}_{\text{FWM}}$ at  $\Delta=0.3$~eV are simpler than
those at the other three values of $\Delta$. 
(3) The oscillations of $\sigma^{xxxx}_{\text{THG}}$ and
$\sigma^{xxxx}_{\text{FWM}}$ are more complicated than those of
$\sigma^{(1);xx}$ and  $\sigma^{xxxx}_{\text{nl}}$. (4) However,
$\sigma^{(1);xx}$ and  $\sigma^{xxxx}_{\text{nl}}$ show the same
oscillatory behavior (the peak position and the period) for $\Delta=0$, 
$0.1$, and $0.3$~eV; they show  different
oscillatory behavior for $\Delta=0.2$~eV. (5) The peak value of each
oscillation increases with the magnetic field. For the magnetic field
we calculated here, they usually
increase by a few times. But for $\sigma^{xxxx}_{\text{THG}}$ at
$\Delta=0$ and $0.1$~eV,  the peak values can increase by about 20
times. (6) At strong
magnetic fields, the magnitude of the conductivities can be close for
different gap parameters, and the strong gap dependence of
conductivities at weak magnetic field becomes unimport-
ant. These features are understood as follows.

The peaks of the conductivity are
induced by the resonant transitions. Based on the energy factor
in Eq.~(\ref{eq:Efactor}), the  resonant conditions can be summarized as
${\cal E}_{s_1s_2}(\epsilon, n, m)=0$, where $\epsilon$ is a
transition energy. Around the resonance, the conductivity behaves as a
Lorentz type divergence $[{\cal E}_{s_1s_2}(\epsilon+\delta E+i\Gamma,
  n, m)]^{-1}$ with respect to $\delta E$.  For GG, we find that only
the interband  transitions lead to resonances, under the condition
\begin{equation}
  \epsilon_{n+m} + \epsilon_n = \epsilon\,. \label{eq:cond}
\end{equation}
Here $n$ is a Landau index varying freely, $(m,\epsilon)$ can be used
to identify channel for resonant optical transition. 
From Eq.~(\ref{eq:cond}) the magnetic field can be solved as $B={\cal B}_e(\epsilon,n,m)$ with
\begin{eqnarray}
  {\cal B}_e&&(\epsilon,n,m)  = \frac{\epsilon^2-(2\Delta)^2}{2e \hbar
    v_F^2} \notag\\
  &&\times \left[(2n+m)
    +2\sqrt{n(n+m)+m^2\left(\frac{\Delta}{\epsilon}\right)^2}\right]^{-1}\,.\quad\label{eq:rescond}
\end{eqnarray}
This solution  exists only for $w>2\Delta$, where the photon energy is
higher than the energy gap and the interband
transitions could be in resonance. The function ${\cal B}_e(\epsilon,n,m)$
decreases with $n$, $m$, and $\Delta$, but increases with
$\epsilon$. For a fixed channel $(m,\epsilon)$, the resonant
transitions induced by the Landau index $n$ occur at the field
${\cal  B}_e(\epsilon,n,m)$.
For large $n$ or $m=0$, $[{\cal B}_e(\epsilon,n,m)]^{-1}\propto n$
indicates that the conductivities are approximately periodic in
$B^{-1}$, similar 
to that of de Haas-van Alphen effect; but for the resonance occurring
between the lowest several Landau levels, the period is also affected by
$m$ and $\Delta/\epsilon$, deviating from the periodicity with
respect to $B^{-1}$. The neighboring resonant peaks in the same
channel occur between the Landau indices $n$ and $n+1$. The period is then
$\Delta B_n={\cal B}_e(\epsilon,n,m)-{\cal B}_e(\epsilon,n+1,m)$. At a large
$n\gg m$, the period $\Delta B_n\approx -\partial_n {\cal
  B}_e(\epsilon,n,m)\approx {\cal B}_e(\epsilon,n,m)/n \propto n^{-2}$ 
decreases quickly with $n$. Therefore at weak magnetic field, the
resonance occurs at large $n$, and it is easier to smear out the
oscillations, as shown in the region for $B<B_c$. 
However, when there exist multiple resonant
channels, their oscillations are mixed and complicated.

We identify the resonant channels. 
For the linear conductivity, there is only one channel as $(m,\epsilon)=(1,\hbar\omega)$. The resonant magnetic field is given by $B_1={\cal
  B}_e(\hbar\omega,n,1)$.
\begin{table}[h]
  \centering
  \begin{tabular}[t]{|c||c|c|c|c|c|c||}
    \hline
    n & 0& 1 & 2 & 3 & 4 & 5\\
    \hline
    $\Delta=0~~$~eV & 68.3&11.7 &6.9  & 4.9& 3.8 & 3.1\\
    $\Delta=0.1$~eV& 22.8 & 6.4 & 3.8 & 2.7 & 2.1 & 1.7\\
    \hline
  \end{tabular}
  \caption{Magnetic field ${\cal B}_e(0.3,n,1)$ for different $n$.}
  \label{tab:magB}
\end{table}
In Table~\ref{tab:magB} we list the values for first several $n$ at
$\Delta=0$ and $0.1$~eV. These values agree with the peak positions
shown in Fig.~\ref{fig:gapCB}~(a). For $\Delta\ge0.15$~eV, there is no
interband resonant transitions and thus no oscillations, as shown in
Fig.~\ref{fig:gapCB}~(a). 

For the nonlinear conductivities, each denominator
includes three energy factors, and there are multiple resonant
channels. In some special conditions, it is possible to have more than
one of these channels occurring simultaneously.  For THG, these
 channels include $(m,\epsilon)=(1,\hbar\omega)$, $(0,2\hbar\omega)$,
 $(2,2\hbar\omega)$, and $(1,3\hbar\omega)$.  They lead to the
 following types of divergences
\begin{equation}
  \begin{array}{rlcl}
    & [{\cal E}_{+-}(\hbar\omega+i0^+,n,1))]^{-1}\,, &&  [ {\cal
        E}_{+-}(2\hbar\omega+i0^+, n, 0)]^{-1} \,,\\
    & [{\cal E}_{+-}(2\hbar\omega+i0^+,n,2)]^{-1}\,,& &    [{\cal E}_{+-}(3\hbar\omega+i0^+, n, 1)]^{-1} \,.
  \end{array}\notag
 \end{equation}
The magnetic fields for these four channels are $B_1={\cal
  B}_e(\hbar\omega,n,1)$, $B_2={\cal 
  B}_e(2\hbar\omega,n,0)$, $B_3={\cal B}_e(2\hbar\omega,n,2)$, and
$B_4={\cal B}_e(3\hbar\omega,n,1)$. For $\Delta=0.3$~eV, only the channel
$B_4$ exists. Because the relevant energy is
$3\hbar\omega$, the resonant levels have larger index 
$n$ than those of the linear conductivity. Inside $[0,10]$~T, the
resonant magnetic fields are $9.0$, $8.2$, $7.4$, $6.8$, $6.3$~T,
$\cdots$ for $n=8$, $9$, $10$, 
$11$, $12$, $\cdots$, respectively. The periods of this channel are
shorter than that of the channel $(m,\epsilon)=(1,\hbar\omega)$ in linear conductivity. For $\Delta=0.2$~eV,
the extra channels $B_2$ and $B_3$ are added, and lead to the 
complicated resonances. The fields for resonant transitions can be
easily identified by using Eq.~(\ref{eq:rescond}) and we do not
present them here. Similar results are found for $\Delta=0.1$~eV and 
$\Delta=0$, where all channels are possible. {Interestingly, at
  weak magnetic field,  these resonant transitions contribute to the
  THG conductivity destructively, 
    where the two-photon resonant transitions contribute with a sign
    opposite to those from one- and three-photon resonant
    transitions. Their cancellation leads to a small THG conductivity
    for cases $\Delta=0.0$ and $0.1$~eV. When the magnetic field
    is strong, the nonequidistant LLs affect the electronic states at
    low energies more than those at high energies. Therefore, the
    magnetic field  affects one-photon resonant transition more than the other two
    transitions, and breaks the cancellation to induce an increment
    up to 20 times for the THG conductivity.}

For NL, the possible resonant channels are
$(m,\epsilon)=(1,\hbar\omega)$, $(0,2\hbar\omega)$, and
$(2,2\hbar\omega)$. They lead to the following divergences
\begin{equation}
  \begin{array}{rlcl}
    & [{\cal E}_{+-}(\hbar\omega + i0^+,n,1)]^{-2}\,,&&[{\cal
        E}_{+-}(\hbar\omega + i0^+,n, 1)]^{-1}\,,\\
    & [{\cal E}_{+-}(\hbar\omega - i0^+,n,1)]^{-1}\,,&& [{\cal
        E}_{+-}(2\hbar\omega+i0^+,n,0)]^{-1} \,,\\
    & [{\cal E}_{+-}(2\hbar\omega+i0^+,n,2)]^{-1}\,.
  \end{array}\notag
\end{equation}
There also exists energy factors like  $(i0^+)^{-1}$ and
$(2\hbar\omega+i0^+)^{-1}$. However, neither of them lead to any
divergence. There exist several resonant channels,
but the one involving ${\cal B}_e(\hbar\omega,n,1)$ leads to
resonances with higher order divergence ($[{\cal E}_{+-}(\hbar\omega + i0^+,n,1)]^{-2}$) and is dominant. Thus at
$\Delta=0$ and $0.1$~eV, although all channels are possible, the
higher order divergence dominates, which is induced by the same resonant
channel as that of linear conductivity. It is not surprising that these
spectra have the same oscillations as those of linear conductivity. For $\Delta=0.2$~eV, the linear
conductivity does not have an available channel, and this higher order
divergence  channel for NL 
is also forbidden. However, for NL the other two channels are
available. For magnetic field in the range of $[0,10]$~T, the resonant magnetic fields for the channel
$(m,\epsilon)=(0,2\hbar\omega)$ are $9.5$, $7.6$, $6.3$,
$5.4$, $4.7$, $4.2$~T, 
$\cdots$ for $n=4$, $5$, $6$, 
$7$, $8$, $9$, $\cdots$, respectively; those for the channel
$(m,\epsilon)=(2,2\hbar\omega)$ are $9.6$, $7.6$, $6.3$,
$5.4$, $4.7$, $4.2$~T, 
$\cdots$ for $n=3$, $4$, $5$, $6$, 
$7$, $8$, $\cdots$, respectively. The period for
$\Delta=0.2$~eV is shorter than those of $\Delta=0$ and $0.1$~eV. For
$\Delta=0.3$~eV, no channel is available, and the conductivity shows no
oscillations. 

For  FWM, two frequencies
$\omega_p$ and $\omega_s$ can result in more channels
$(m,\epsilon)=(1,\hbar\omega_s)$, $(1,\hbar\omega_p)$,
$(0,\epsilon_a)$, $(2,\epsilon_a)$, $(0,2\hbar\omega_p)$,
$(2,2\hbar\omega_p)$, and $(1,\epsilon_b)$ with $\epsilon_a=\hbar\omega_p-\hbar\omega_s$ and
$\epsilon_b=2\hbar\omega_p-\hbar\omega_s$.  They lead to the following divergences 
\begin{equation}
  \begin{array}{clcl}
    &[{\cal E}_{+-}(\hbar\omega_s- i0^+,n,1]^{-1}\,, & &
    [{\cal E}_{+-}(\hbar\omega_p + i0^+,n,1)]^{-1}\,,\\
    & [{\cal E}_{+-}(\epsilon_a+i0^+, n, 0)]^{-1} \,, & & [{\cal E}_{+-}(\epsilon_a+i0^+,n,2)]^{-1}\,, \\
    &  [{\cal E}_{+-}(2\hbar\omega_p+i0^+,n,0)]^{-1}\,, && [{\cal E}_{+-}(2\hbar\omega_p+i0^+,n,2]^{-1}\,,\\
    &[ {\cal E}_{+-}(\epsilon_b,n,1)]^{-1}\,.
  \end{array}\notag
\end{equation}
In the limit $\omega_s=\omega_p$, the FWM 
conductivity is reduced to that of the NL conductivity, and the higher
order divergence $(\delta E+i0^+)^{-2}$ is a combination of two energy
factors. We take the case $\Delta=0.3$~eV as an example. The resonances can be induced by the
energy factors ${\cal E}_{+-}(2\hbar\omega_p+i0^+,n,0)$, ${\cal
  E}_{+-}(2\hbar\omega_p+i0^+,n,0)$, and ${\cal
  E}_{+-}(\epsilon_b+i0^+,n,1)]^{-1}$. For 
  $B\in[0,10]$~T, the resonant magnetic field  
  ${\cal B}_e(\epsilon_b,n,1)$ is $9.9$, $7.1$, $5.5$, and $4.5$ for $n=2$,
  $3$, $4$, and $5$, respectively. These field values determine the main
  peaks. The fields ${\cal 
    B}_e(2\hbar\omega_p,n,0)$ and ${\cal B}_e(2\hbar\omega_p,n-1,2)$ are
  very close with values around $9.35$, $8.68$, $8.1$,
  and $7.6$ for $n=13$, $14$, $15$, and
  $16$, respectively. These field values determine the small changes
  on both sides of the main peaks. Other peaks can be understood in a
  similar fashion. 

\subsection{Results of DG}
\begin{widetext}
\begin{figure*}[ht]
  \centering
  \includegraphics[width=6cm]{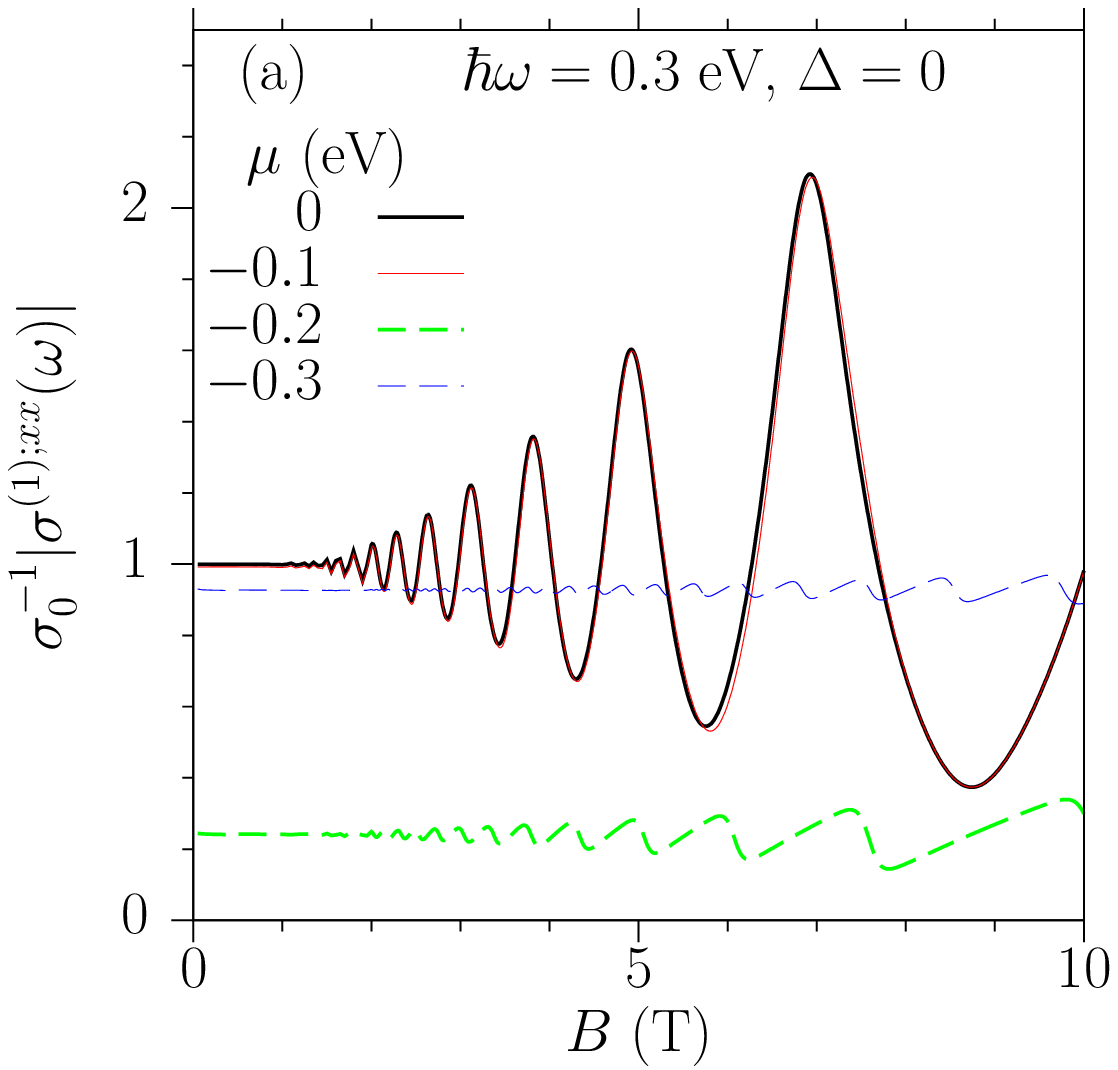}~~
  \includegraphics[width=6cm]{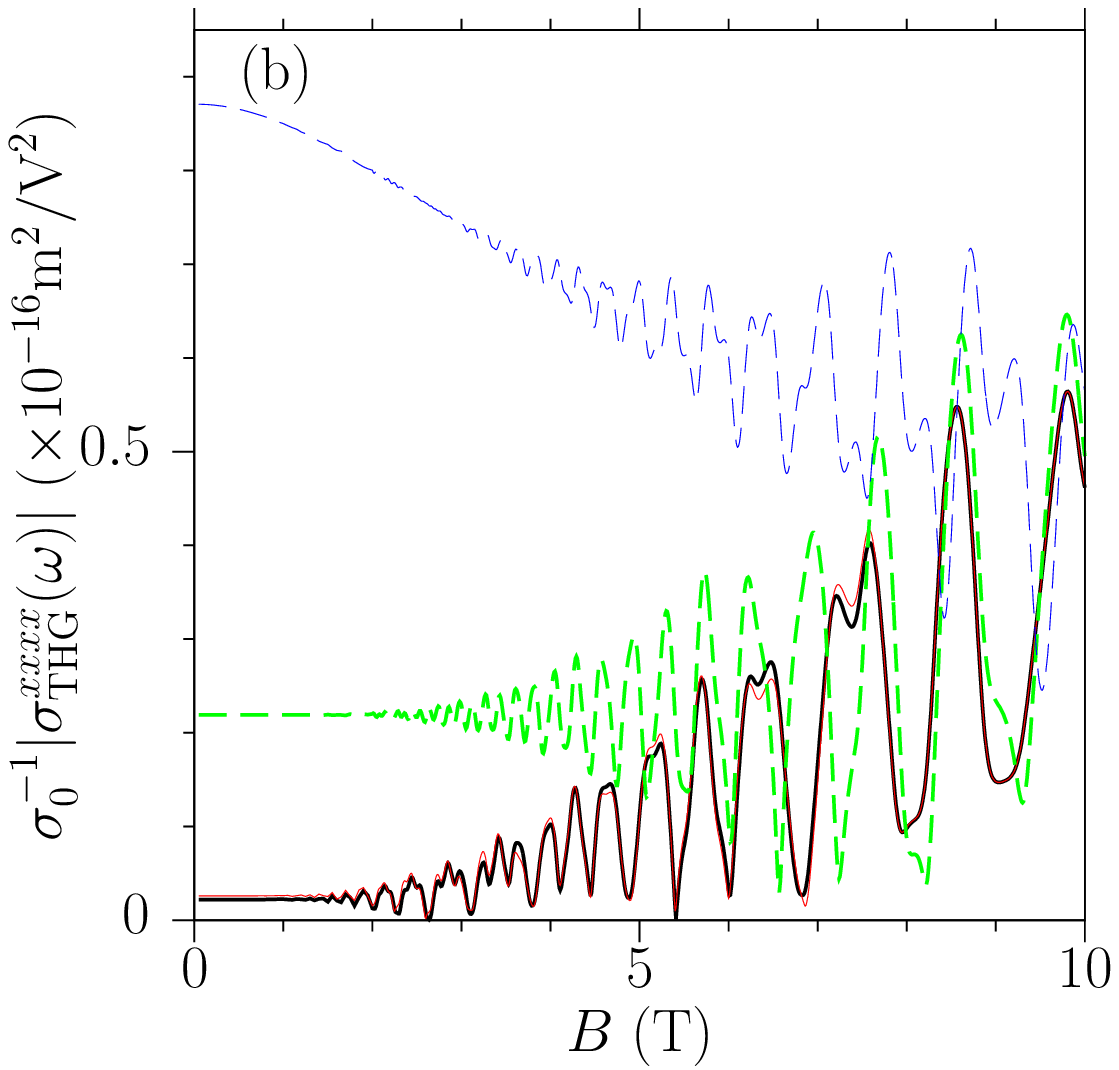}\\
  \includegraphics[width=6cm]{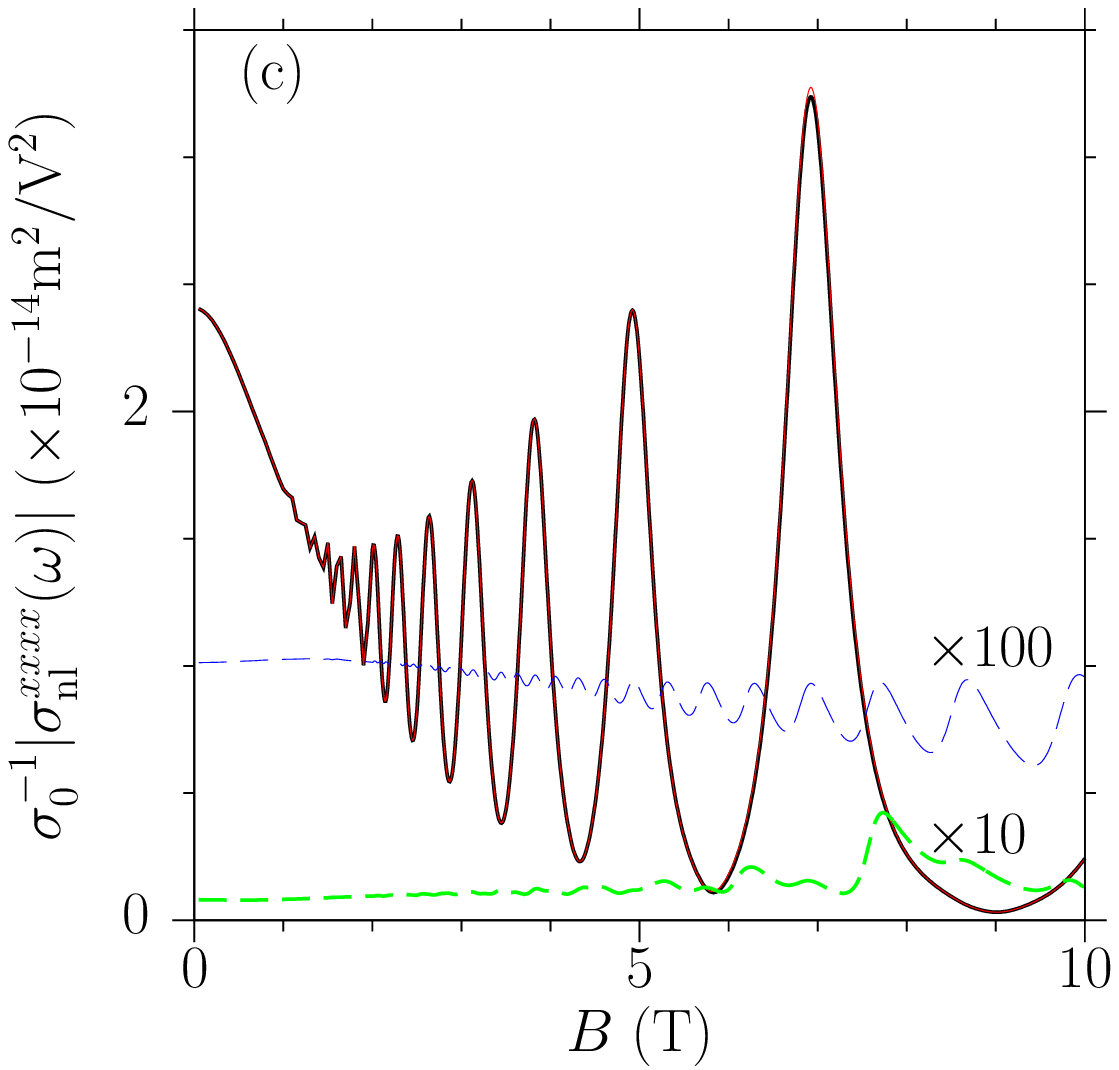}~~
  \includegraphics[width=6cm]{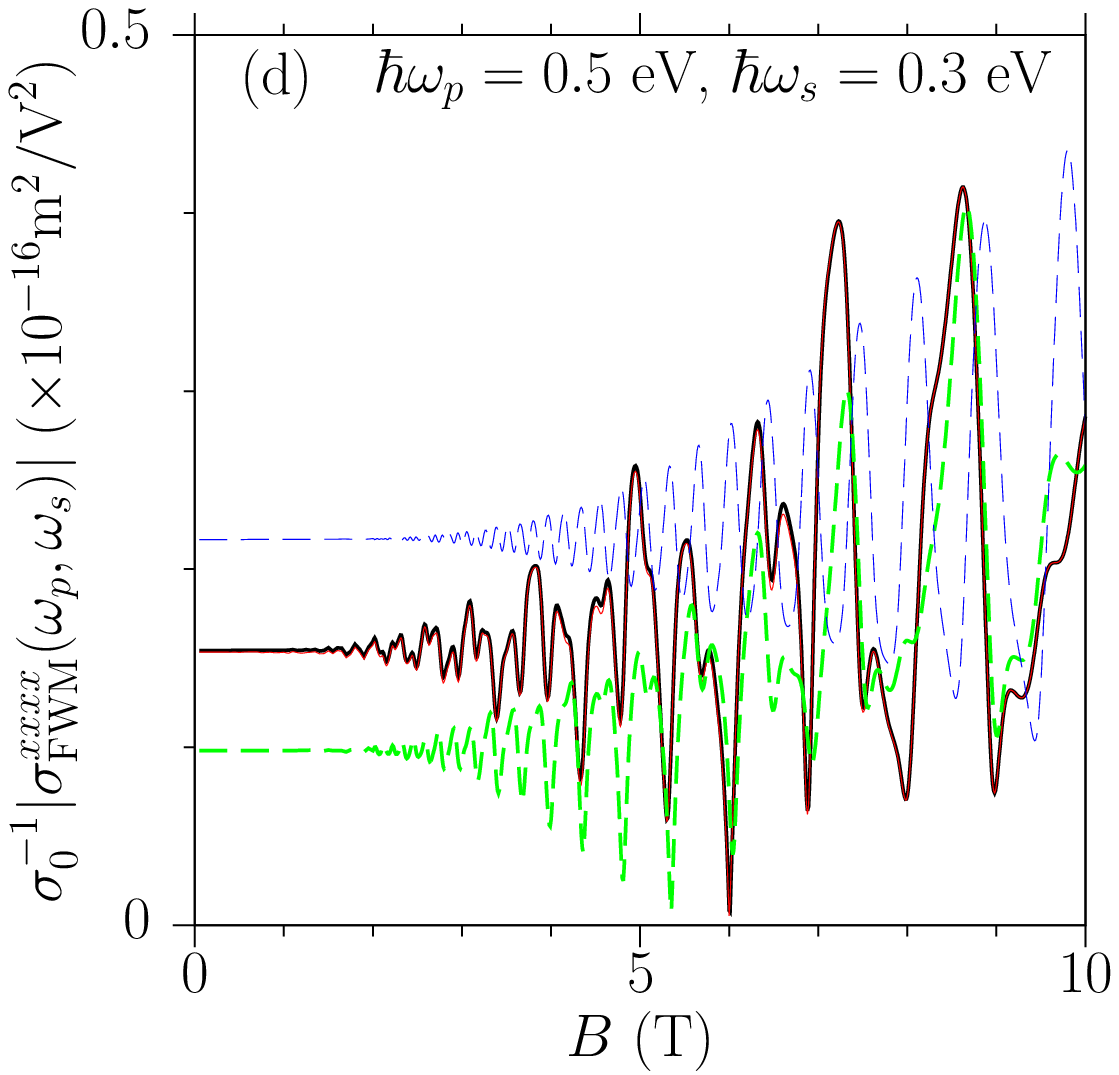}
  \caption{(color online) Magnetic field dependence of the absolute values of optical
    conductivities of  DG (a) $|\sigma^{(1);xx}(\omega)|$, (b)
    $|\sigma^{xxxx}_{\text{THG}}(\omega)|$,
    (c) $|\sigma^{xxxx}_{\text{nl}}(\omega)|$,
    (d) $|\sigma^{xxxx}_{\text{FWM}}(\omega_p,\omega_s)|$. The
    parameters are $\hbar\omega=\hbar\omega_s=0.3$~eV,
    $\hbar\omega_p=0.5$~eV, $\Gamma=10$~meV, $\Delta=0$, and $T=10$~K. In
    (c), the curves for $\mu=-0.2$ and $-0.3$~eV are scaled by 10 and 100
     times, respectively.}
  \label{fig:dopeCB}
\end{figure*}
\end{widetext}

In  Fig.~\ref{fig:dopeCB}, we present the absolute values of
conductivities $\sigma^{(1);xx}(\omega)$, $\sigma^{xxxx}_{\text{nl}}(\omega)$,
$\sigma^{xxxx}_{\text{THG}}(\omega)$, and
$\sigma^{xxxx}_{\text{FWM}}(\omega_p,\omega_s)$ 
of DG at the chemical potential $\mu=0$, $-0.1$, $-0.2$, and $-0.3$~eV, by varying the
magnetic field $B$ from $0$ to $10$~T.  Because of the finite doping,  the transverse
conductivity components $\sigma^{(1);xy}$, 
$\sigma^{(3);xyxx}$, $\sigma^{(3);xxyx}$, and
$\sigma^{(3);xxxy}$ are nonzero.

In the limit of zero magnetic field, the chemical  potential
dependence of the conductivities has been discussed in
literature and also  shown in Fig.~\ref{fig:spectraB}.
The effect of the chemical potential is very similar to an energy gap,
but they also have essential differences. With
increasing the magnetic field, the conductivities at all chemical
potentials oscillate. Compared to the
conductivities of GG, the optical response of DG shows some different
magnetic field dependence: (1) As opposed to the gap parameter
dependence, the conductivities have little difference for $\mu=0$ and
$-0.1$~eV. All the photon energies we calculated are
away from resonance for these two chemical potentials. (2) It can be
seen that for a given resonant channel the oscillation period does not
depend on the chemical potential. (3) At $\mu=-0.2$ and $-0.3$~eV,
$\sigma^{(1);xx}$ and $\sigma^{xxxx}_{\text{nl}}$ show additional
fluctuations. They are induced by the difference between the left and right
circularly polarized components. 
 In the dc limit,  $\sigma^{(1);xy}$ is induced
by quantum Hall effects, and its value is determined by the number of
doped LLs and shows one plateau for each LL. Similar natures also exist in
the longitudinal components and lead to such plateau like
fluctuations. 

In contrast to the undoped case in GG, where the resonance occur only
between the interband transitions, here resonant intraband
transitions are 
also possible.  For DG, the values for all possible  magnetic fields 
are given by
\begin{eqnarray}
  {\cal B}_s(\epsilon,n,m) &=& \frac{\epsilon^2}{2e\hbar v_F^2}
  \frac{1}{(\sqrt{n+m}+s\sqrt{n})^2}\notag\\
  &\approx& 760 \left(\frac{\epsilon}{\text{eV}}\right)^2\frac{1}{(\sqrt{n+m}+s\sqrt{n})^2}\,,
\end{eqnarray}
with $s=-1$ for resonant intraband transitions and $s=1$ for resonant
interband transitions. A simple calculation shows that there is no
resonant  intraband transition for the  photon frequencies considered
and the magnetic field in the range of $[0,10]$~T. Thus only the
resonant interband transitions are possible in our parameters. The
chemical potential does not affect the 
value of the resonant magnetic field, but it can block some channels for a
number of Landau index $n$.  
This can be clearly seen from the conductivity $\sigma^{(1);xx}(\omega)$
at $\mu=0$, $-0.1$, and
$-0.2$. The energy of involved
electron/hole in the resonant transition can be approximated as
$\pm\hbar\omega/2\sim \pm 0.15$~eV. At room temperature $T=300$~K, both
the electron and hole are almost empty for the case $\mu=-0.3$~eV,
which lead to 
\begin{widetext}
 \begin{figure*}[ht]
  \centering
  \includegraphics[width=6cm]{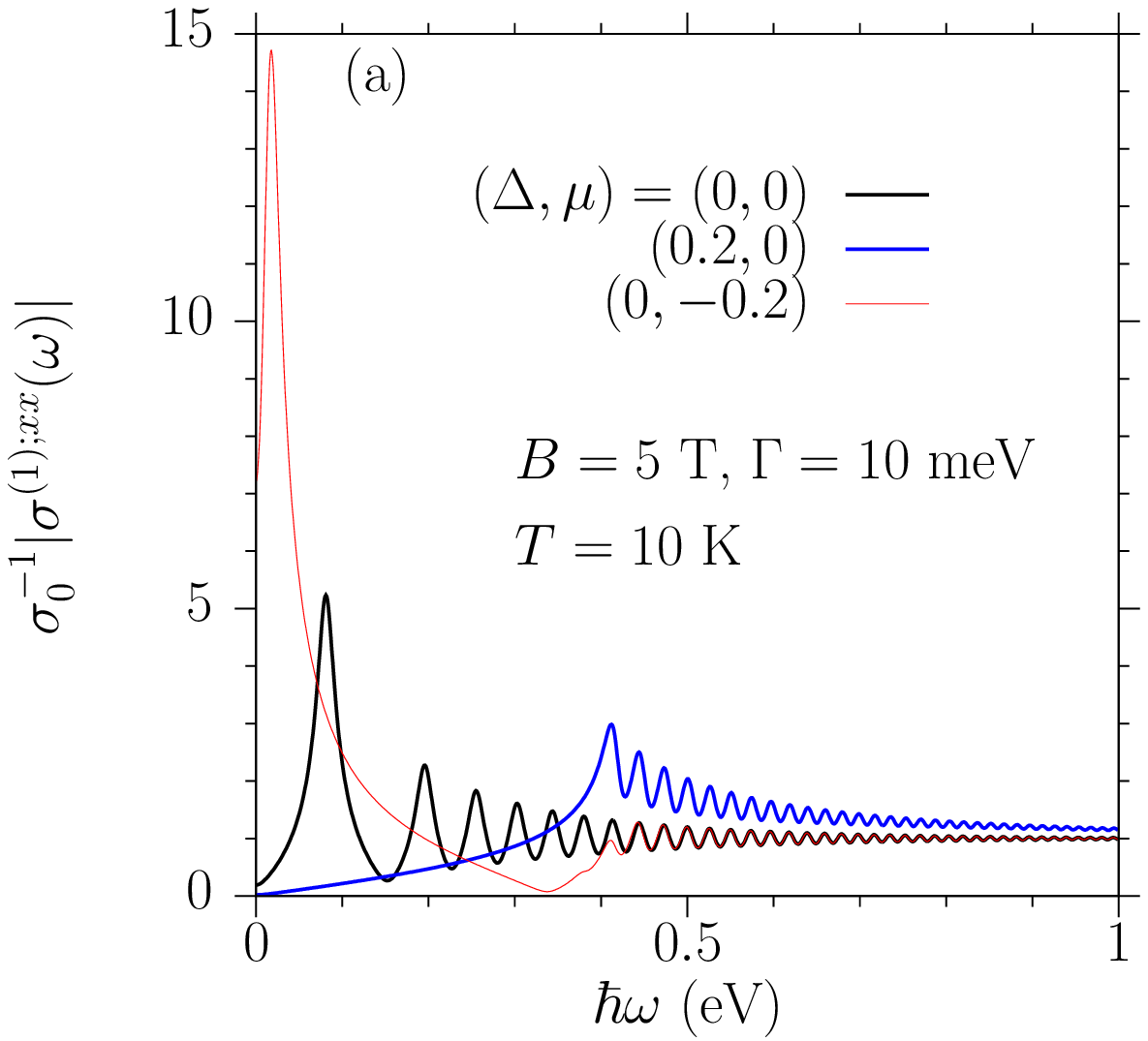}~~
  \includegraphics[width=6cm]{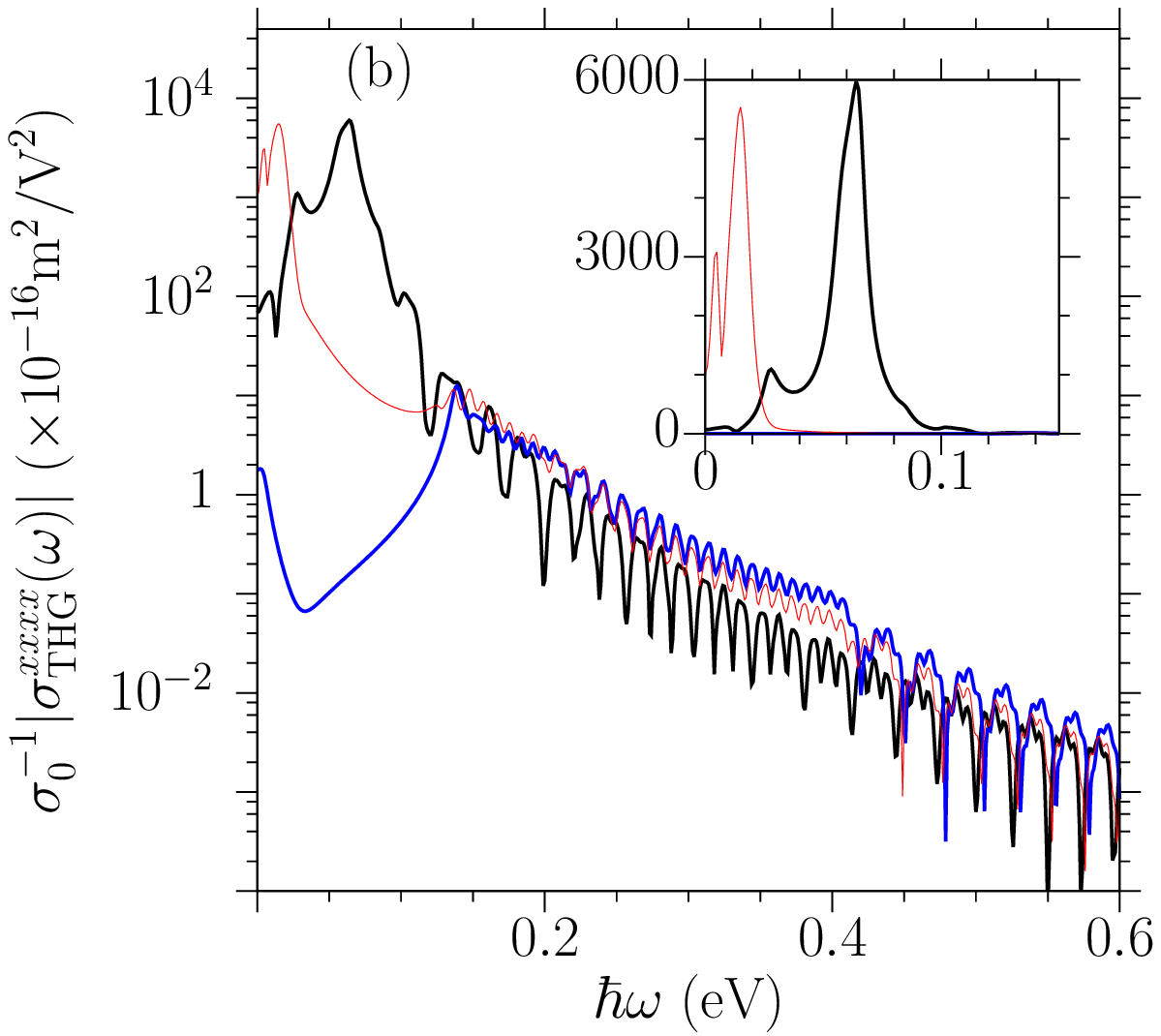}\\
  \includegraphics[width=6cm]{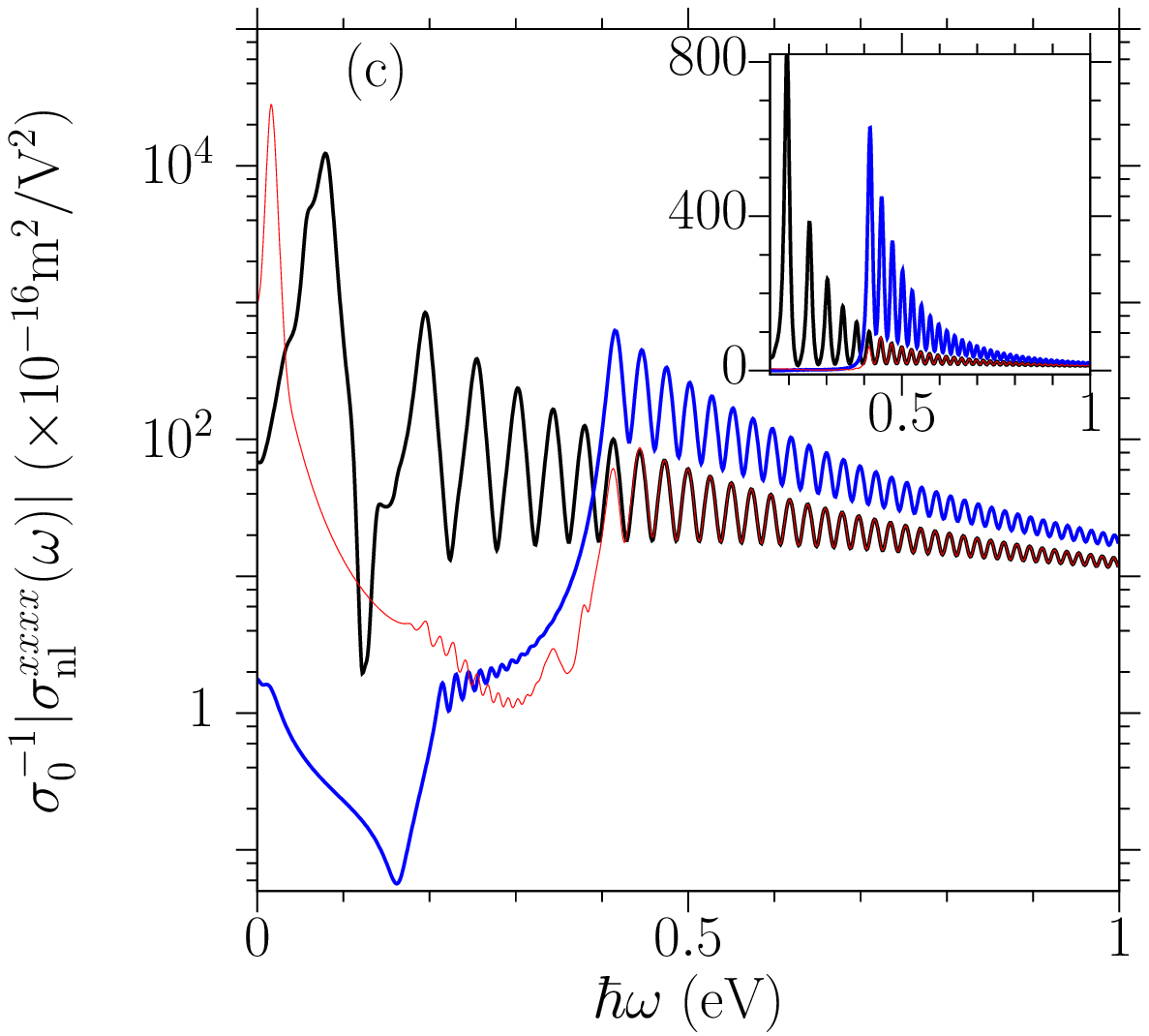}~~
  \includegraphics[width=6cm]{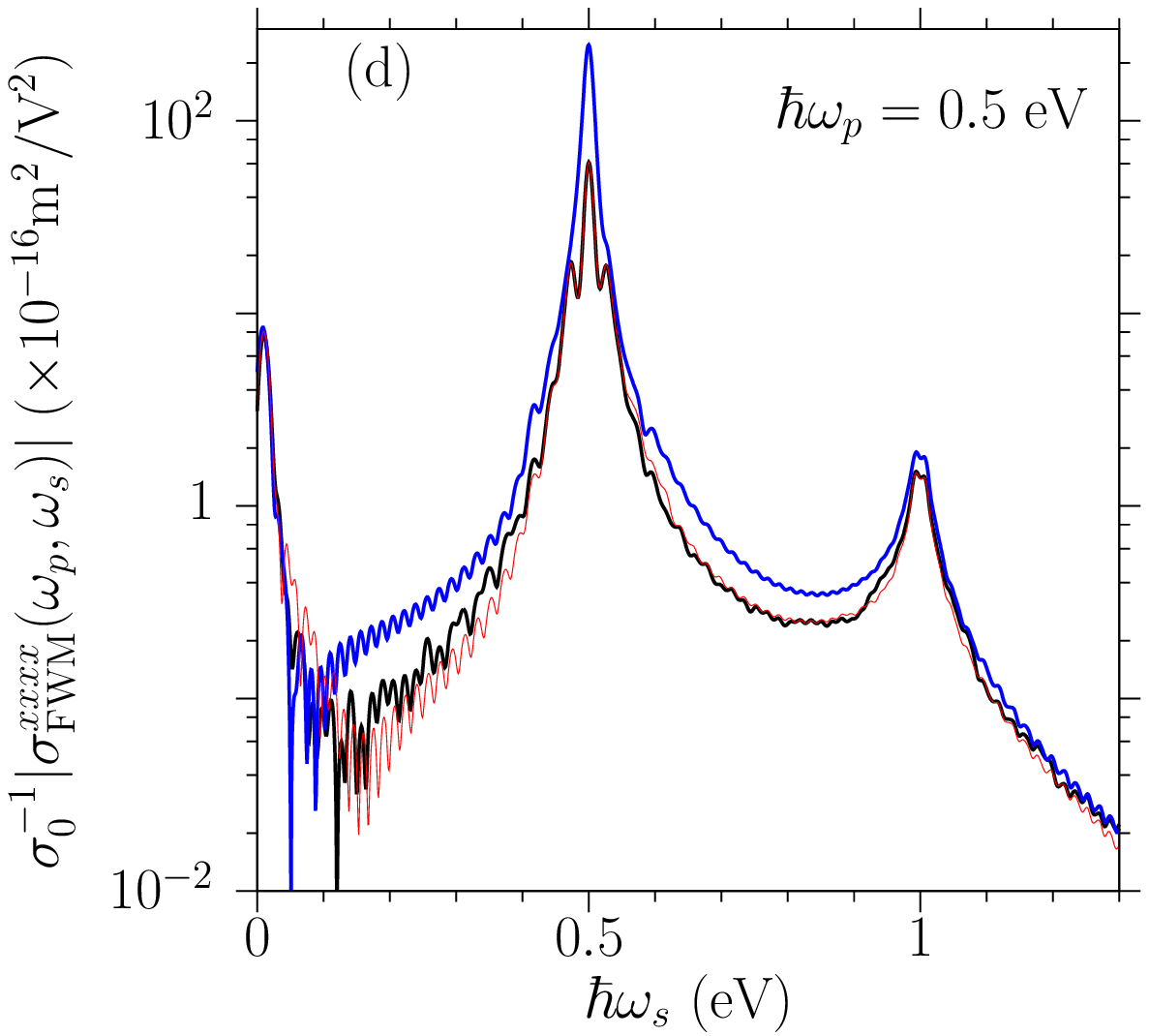}
  \caption{(color online) Spectra of absolute values of optical
    conductivities of GG  and DG at $B=5$~T for different chemical
    potential and gap 
    parameter $(\Delta,\mu)$. (a) $|\sigma^{(1);xx}(\omega)|$, (b) 
    $|\sigma^{xxxx}_{\text{THG}}(\omega)|$,
    (c) $|\sigma^{xxxx}_{\text{nl}}(\omega)|$,
    (d) $|\sigma^{xxxx}_{\text{FWM}}(\omega_p,\omega_s)|$. The gap
    parameter and the chemical potential are $(\Delta,\mu)=(0,0)$,
    $(0.2,0)$, and $(0,-0.2)$~eV. Other
    parameters are $\hbar\omega_p=0.5$~eV, $\Gamma=10$~meV, and
    $T=10$~K. The inset in (b) gives the details at low photon energy
    $\hbar\omega\in[0,0.15]$~eV, and that in (c) gives the details at
    high photon energy 
    $\hbar\omega\in[0.15,1]$~eV; both $y$-axes are in linear scale. The
    $y$-axes of the
    main diagrams of (b), (c), and (d) are in logarithmic scale.}
  \label{fig:spectraStrongB}
\end{figure*}
\end{widetext}
the vanishing  of the resonant interband transition. Similar
results can be found for $\sigma^{xxxx}_{\text{nl}}(\omega)$. For the
nonlinear conductivities, the occupation of levels only affects the
transition involving $w_3$ in Fig.~\ref{fig:illusnontrans}. Because
some channels are determined by more than one resonant transition, it is
not a surprise that the resonant peaks of $\sigma^{xxxx}_{\text{THG}}$ and
$\sigma^{xxxx}_{\text{FWM}}$ are so complicated.

 \section{Optical conductivities at strong magnetic field \label{sec:spectraStrongB}}
In this section we present the conductivities at a 
strong magnetic field $B=5$~T. The cyclotron energy is 
$\hbar\omega_c=81$~meV, from which the lowest several energy levels 
are $\epsilon_n=0$, $81$, $114$, $140$, $162$, $181$~meV at
$n=0,1,2,3,4,5$ for $\Delta=0$ and $\epsilon_n=0.2$, $0.216$, $0.230$,
$0.244$, $0.257$ at $n=0,1,2,3,4,5$ for $\Delta=0.2$. Both cases show
explicit discrete levels. Our numerical results are shown in
Fig.~\ref{fig:spectraStrongB} for $\sigma^{(1);xx}(\omega)$, 
$\sigma^{xxxx}_{\text{THG}}(\omega)$, $\sigma^{xxxx}_{\text{nl}}(\omega)$, and
$\sigma^{xxxx}_{\text{FWM}}(\omega_p,\omega_s)$ at three different
chemical potentials and gap parameters $(\Delta,\mu)=(0,0)$, $(0.2,0)$, and
$(0,-0.2)$~eV, which are noted as intrinsic graphene (IG), GG, and DG
in the following.
 
The linear conductivity is plotted in Fig.~\ref{fig:spectraStrongB}
(a). All spectra show oscillations. For photon energies
$\hbar\omega>0.4$~eV, the oscillations are similar for all three
cases. With the increase of the photon energy, the oscillation period
and amplitude decrease. From our results in previous
sections, we understand these oscillations from the resonant interband
transitions for $\hbar\omega>E_g$, which occur between the electronic
state ``$sn$''$=$``$-n$'' and the state ``$+ (n+1)$'' or between 
``$- (n+1)$'' and  ``$+ n$''. The peak positions are
determined by the transition energy $\hbar\omega = \epsilon_{n+1} +
\epsilon_n$, with $n\ge0$ for GG and $n\ge6$ for IG and
DG. In this  region, the conductivity oscillates around the value 
of conductivity at zero magnetic field. Furthermore, the cases for IG
and DG are almost identical, because
they have the same optical resonant transitions between the same
LLs. When the photon energy $\hbar\omega<0.4$~eV, the conductivities show
different behavior. For the case of IG, the
interband resonant transitions can be extended to $n=1$; while the
special case at $n=0$,  corresponding to a transition energy
$\hbar\omega=81$~meV, includes both intraband and
interband transitions. For the case of DG, the
interband transitions are blocked; however, there is one extra peak at
$18$~meV. This peak results from the intraband transition between LLs
of ``$- (n+1)$'' and ``$- n$'' at $n=6$. It is the modified
Drude contribution for LLs.  For the case of GG, all interband
transitions are forbidden,  and they result in a smooth linear conductivity.
The spectrum of $\sigma^{xxxx}_{\text{nl}}(\omega)$ in
Fig.~\ref{fig:spectraStrongB} (c) shows  dependence on
photon energy similar to that of the linear conductivity.

The spectra of THG conductivity 
$\sigma^{xxxx}_{\text{THG}}(\omega)$ are shown in Fig.~\ref{fig:spectraStrongB}
(b). For $\hbar\omega$ less than about $0.15$~eV, the spectra of the conductivity
 include a few peaks for IG and DG, and the values can be as
large as $5\times 10^{-13} \sigma_0 \text{V}^2/\text{m}^2$. The peaks locate at around $8$, $27$, and
$64$~meV for IG, and $4$ and $14.5$~meV for DG. By checking the energies
for interband resonant transitions and possible intraband
resonant transitions,  the peak at $64$~meV is induced by the
three-photon resonant interband transitions between ``$- 1$'' and
``$+ 2$'' or between ``$- 2$'' and ``$+ 1$''; the peaks at $27$~meV or
$8$~meV are induced by
resonant intraband transitions, between the LLs around the Fermi
surface. From the illustration diagrams listed in
Fig.~\ref{fig:illusnontrans}, the resonant transitions can occur at any
stage of the three transitions; some of them have no requirement for
the occupation of the initial and final states. However, the dominant
contributions are still from the resonant transitions from an occupied
state to an empty state. A similar analysis can be applied to the case of
DG, where the first two peaks comes from the intraband transitions
around $n=6$. For higher photon energies, many oscillations appear; 
 they are induced by interband transitions, as discussed in previous
section. Even at very high photon energies, the conductivity can be
tuned by about two orders of magnitude in one oscillation. {Because LLs are nonequidistant, the resonant transitions at high
    photon energies involve very high LL indices, where the magnetic
    field only modulates the original band structure
    slightly. Therefore, similar to that of graphene without the
    presence of a magnetic field, the conductivity decreases as power law of
    frequency for high photon energies \cite{NewJ.Phys._16_53014_2014_Cheng}, but modulated by 
    oscillations from the LLs. }

The spectra of the conductivity of FWM  process
$\sigma^{xxxx}_{\text{FWM}}(\omega_p,\omega_s)$ are plotted in
Fig.~\ref{fig:spectraStrongB} (d). It keeps the features of the
results at zero magnetic field, and it shows three resonant peaks at
$\hbar\omega_s=0$, $\hbar\omega_p$, and $2\hbar\omega_p$. The first is induced by the Drude-like contributions. At the second
one, the FWM process is reduced to NL process. Because the one-photon
absorption exists in our parameters, the two-photon absorption process
diverges, and the conductivity exhibits  a divergent peak. At the third one, the FWM process corresponds to a current
injection process, which diverges too. However, although the magnetic
field modulates the conductivity, the changes are smaller than those in
the conductivities for THG and NL processes.

Here we note that the resonant intraband transitions give huge
responses, similar to their contributions at zero magnetic
field. However, the magnetic field brings an advantage for
controlling the position of resonant peak, from $\hbar\omega=0$ at zero
magnetic field to a nonzero value for nonzero magnetic field.
 
\section{Conclusion \label{sec:conclusion}} 
In this work, we have investigated the perturbative linear and
third order conductivities of gapped graphene and doped
graphene in the presence of a perpendicular magnetic field. The electron
dynamics are solved from the equation of motion using Landau
levels. The light matter interaction is described in the length gauge and
the scattering is included in a phenomenological way with only one
relaxation parameter. We discuss the nonlinear processes for third harmonic
generation, the nonlinear corrections to 
the linear conductivity, and four wave mixing.

We first show that the Landau levels form a good
basis functions even at very weak magnetic field, which can be used
to calculate the conductivities of systems without magnetic field. We
apply this approach to a doped graphene and the results agree with the
calculation from an analytic expression in literature. Using the same
approach, we present the optical
conductivities of a gapped graphene. 
Similar to the chemical potential related resonant transition in doped
graphene, there exist energy gap related resonant transitions,
occurring when any of the involved photon energies matches the band gap. 
The main difference lies in the absence of the Drude contribution in a 
gapped graphene, which leads to its temperature insensitivity and
dielectric nature below the band gap.  

At strong magnetic fields, the Landau
levels are discrete and there exist many resonant transitions. Varying
the magnetic field in the range of $[0,10]$~T, the nonlinear 
conductivity can be tuned up to $1-2$ orders of
magnitude. There exist different resonant channels, especially in the
nonlinear optical response. Some of them can be turned on and off by
tuning the the gap  
parameter and the chemical potential, leading to different oscillation
features. We also present a simple condition to identify these resonant transitions. We calculate the spectra of these 
conductivities at strong magnetic field. The spectra show oscillations, which are induced by the resonant transitions between
different Landau levels. At small photon energies, the conductivity of
DG show peaks due to the resonant intraband transitions, corresponding
to a modified Drude contribution.

In our calculations, the phenomenological relaxation time
approximation is a very rough treatment, which is intended to describe all
microscopic relaxation processes, many-body effects, and thermal
effects in just one parameter. For optical properties in most materials,
such treatment will not lead to difficulties because the time scale
of optical field is much faster than the scattering processes, and
such a treatment is mostly used to remove the divergence in the
calculation. Although it is not clear whether or not this is still the
case for graphene in a strong magnetic field, the calculation
presented here can at least indicates interesting qualitative behavior
at the considered frequencies and field strengths, such as the
oscillations and the dependence on the gap parameter and chemical
potentials.  These  properties are likely to remain in more
sophisticated calculations. Because all these oscillations can be
tuned by the strength of the magnetic field, these calculations indicate a
new way to control the optical 
response in the terahertz to the far-infrared.

{To connect with experiments like four wave mixing and third
harmonic generation, it might be convenient to estimate  the
output intensity from the input ones. For 
free-standing graphene, the radiated electric field can be
calculated \cite{Phys.Rev.Lett._108_255503_2012_Yao,boyd_nonlinearoptics} by
$E^x(2\omega_1+\omega_2)| \approx\sigma^{(3);xxxx}(\omega_1,\omega_1,\omega_2)/(2c\epsilon_0)[E^x(\omega_1)]^2E^x(\omega_2)$,
and then the output intensity $I(2\omega_1+\omega_2)$ is given by
\begin{eqnarray}
  I(2\omega_1+\omega_2) &\approx&
  \frac{|\sigma^{(3);xxxx}(\omega_1,\omega_1,\omega_2)|^2}{(2c\epsilon_0)^{4}}
  [I(\omega_1)]^2[I(\omega_2)] \notag\\
  &\approx& 46.6
  \left|\frac{\sigma^{(3);xxxx}(\omega_1,\omega_1,\omega_2)}{10^{-14}\sigma_0}\right|^2\notag\\
  && \times \left[\frac{I(\omega_1)}{1\text{GW/m}^2}\right]^2\left[\frac{I(\omega_2)}{1\text{GW/m}^2}\right]\,.
\end{eqnarray}
The output intensity is directly determined by the square of the
conductivity. For fixed input laser pulse, the maximal output
intensity can be found when the conductivities are maximized by
tuning the magnetic field and the gap parameter/chemical
potential. If the graphene is inside a complicated structure, the
output can be strongly modified by the design of the structure. }

\acknowledgments
This work has been supported by CIOMP Y63032G160, QYZDB-SSW-SYS038
of Chinese Academy of Sciences, and NSFC 11774340. J.L.C acknowledges valuable
discussions with Prof. J.E. Sipe. 

\appendix

\section{Matrix elements of position and velocity operators \label{app:v}}
Because there is no coupling between different valleys, the position
and velocity operators only have matrix elements between the
electronic states in the same valley. The matrix elements of position
operators are given as
\begin{eqnarray}
  &&\int d\bm r \Psi_{\nu s_1n_1k_1}^{\dag}(\bm r) x \Psi_{\nu
    s_2n_2k_2}(\bm r)\notag\\
  &&=
  \delta(k_1-k_2)\left(\xi_{\nu ;s_1n_1,s_2n_2}^x-
  l_c^2 k_2 \delta_{s_1,s_2}\delta_{n_1,n_2}\right)\,,\\
&&  \int d\bm r \Psi_{\nu s_1n_1k_1}^{\dag}(\bm r) y \Psi_{\nu
    s_2n_2k_2}(\bm r) \notag\\
  &&=
  \delta(k_1-k_2)\xi_{\nu;s_1n_1,s_2n_2}^y -i\frac{\partial \delta(k_1-k_2)}{\partial k_2}\delta_{s_1,s_2}  \delta_{n_1,n_2}\,,\quad
\end{eqnarray}
with
\begin{eqnarray*}
  \xi_{\nu;s_1n_1,s_2n_2}^x &=& \int dx\Phi_{\nu s_1n_1}^{\dag}(x)x\Phi_{\nu s_2n_2}(x)\,,\\
  \xi_{\nu;s_1n_1,s_2n_2}^y&=&-\frac{l_c^2}{\hbar}\int dx
\Phi_{\nu s_1n_1}^{\dag}(x)p_x \Phi_{\nu s_2n_2}(x)\,.
\end{eqnarray*}
Further the circularly polarized components are
\begin{eqnarray}
  \xi_{\nu;s_1n_1,s_2n_2}^+&=&-il_c\int
  dx\Phi_{\nu s_1n_1}^{\dag}(x)\hat a^\dag\Phi_{\nu s_2n_2}(x)\,,\notag\\
  \xi_{\nu;s_1n_1,s_2n_2}^-&=&~~il_c\int
  dx\Phi_{\nu s_1n_1}^{\dag}(x)\hat a\Phi_{\nu s_2n_2}(x)\notag\\
  &=&(\xi_{\nu;s_2n_2,s_1n_1}^+)^\ast\,. \label{eq:xiconj}
\end{eqnarray}
Using the properties of the operator in Eq.~(\ref{eq:operator}), we
can get Eq.~(\ref{eq:bc}).   Using the symmetry between two valleys in
Eq.~(\ref{eq:valleyeq}), we verify 
\begin{equation}
  \xi^\tau_{\nu;s_1n_1,s_2n_2} =s_1s_2\xi^\tau_{\overline{\nu};\overline{s}_1n_1,\overline{s}_2n_2}\,.\label{eq:xivalley}
\end{equation}
Therefore, the values of all $\xi^\tau_{\nu;s_1n_1,s_2n_2}$ can be
generated from $\xi^+_{+;s_1(n+1),s_2n}$ by Eqs.~(\ref{eq:xiconj}) and
(\ref{eq:xivalley}).

The velocity operator is $\bm v_\nu=[\bm r,H_{\nu;p+e\bm A(\bm
  r)}]/(i\hbar)$. Because the level energy is independent of $k$, it
is calculated directly to give
\begin{equation*}
\int d\bm r \Psi_{\nu s_1n_1k_1}^{\dag}(\bm r) \bm v_\nu \Psi_{\nu s_2n_2k_2}(\bm r)
= \delta(k_1-k_2) \bm v_{\nu ;s_1n_1,s_2n_2}\,,
\end{equation*}
with
\begin{equation*}
\bm v_{\nu;s_1n_1.s_2n_2} = i\hbar^{-1}(s_1\epsilon_{n_1}-s_2\epsilon_{n_2}) \bm \xi_{\nu;s_1n_1,s_2n_2}\,.
\end{equation*}

\begin{figure}[tph]
  \centering
\includegraphics[width=8cm]{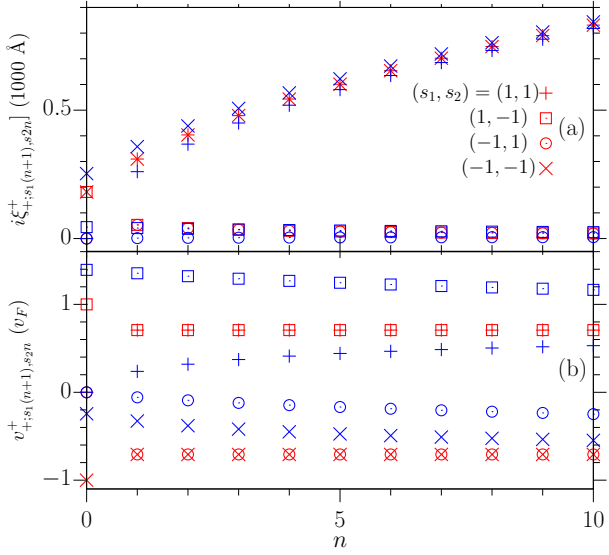}
\caption{(color online) The landau level index dependence of (a)
  $i\xi_{+;s_1(n+1),s_2n}^+$ and (b) $v_{+;s_1(n+1),s_2n}^+$ for
  $B=1$~T and $\Delta=0$ (red symbols) and $\Delta=0.1$~eV (blue
  symbols).} 
\label{fig:xienvp}
\end{figure}

In Fig.~\ref{fig:xienvp} we give $i\xi_{+;s_1(n+1),s_2n}^+$ and
$v_{+;s_1(n+1),s_2n}^+$ for different $n$ and $s_i$. For the special
term $n=0$, we have $\xi^+_{+;s_11,s_20}=-i l_c \sqrt{1-s_1\alpha_1}/\sqrt{2}
\delta_{s_2,-1}$. The optical dipole matrix elements between
the same bands ($s_1s_2=1$) have larger values which increase with $n$. The interband matrix elements
($s_1s_2=-1$) have smaller values: $\xi_{+;+(n+1),-n}^+$
decreases with $n$; $\xi_{+,-(n+1),+n}^+$ first increases then
decreases, where there exists a maximum value depending on the ratio
$\Delta/\hbar\omega_c$. For the velocity matrix elements, both intraband
and interband terms have similar amplitude. For graphene, they have the values
$v_{+;+(n+1),sn}^+=-v_{+;-(n+1),sn}^+=v_F/\sqrt{2}$ for all $n\ge1$ and
$v_{+;s_11,s_20}^+=s_1v_F\delta_{s_2,-1}$. 

\section{Perturbative optical conductivities\label{app:j}}
For a weak electric field, Eq.~(\ref{eq:ksbe}) can be solved
perturbatively by expanding $\rho_\nu(t)$ up to the third order of
the electric field as 
\begin{eqnarray}
  \rho_{\nu}(t) &=& \rho^0_{\nu} + \int\frac{d\omega_3}{2\pi} \widetilde{\cal
    P}^{(1);\gamma}_{\nu}(w_3) E^{\overline{\gamma}}({\omega_3})
  e^{-i\omega_3 t} \notag\\
  &+&
  \int\frac{d\omega_2 d\omega_3}{(2\pi)^2} \widetilde{\cal
    P}_{\nu} ^{(2);\beta\gamma}(w_0,w_3) E^{\overline{\beta}}({\omega_2})E^{\overline{\gamma}}({\omega_3})
  e^{-i(\omega_2+\omega_3) t}  \notag\\
  &+&  \int\frac{d\omega_1d\omega_2 d\omega_3}{(2\pi)^3} \widetilde{\cal
    P}_{\nu}
  ^{(3);\alpha\beta\gamma}(w,w_0,w_3)E^{\overline{\alpha}}({\omega_1})
  E^{\overline{\beta}}({\omega_2})\notag\\
  &&\times E^{\overline{\gamma}}({\omega_3}) e^{-i(\omega_1+\omega_2+\omega_3) t} +\cdots 
\end{eqnarray}
where $E^{\alpha}({\omega})=\int dt E^{\alpha}(t) e^{i\omega
  t}$ is the Fourier transform of the field $E^\alpha(t)$,
$w_3=\hbar\omega_3+i\Gamma$, $w_0=\hbar(\omega_2+\omega_3)+i\Gamma$,
$w=\hbar(\omega_1+\omega_2+\omega_3)+i\Gamma$; the dependence on
$w_3$, $w_0$, and $w$ is clear from the following
expressions. Substituting the expansion above into Eq.~(\ref{eq:ksbe})
and comparing the terms with the same order of electric field at both
sides, we get  
\begin{eqnarray}
  \widetilde{\cal P}_{\nu;s_1n_1,s_2n_2}^{(1);\gamma}(w_3) &=&
  \frac{e\xi^\gamma_{\nu;s_1n_1,s_2n_2}(f_{s_2n_2}-f_{s_1n_1})}{w_3-(s_1\epsilon_{n_1}-s_2\epsilon_{n_2})}\,,\label{eq:p1}\\
  \widetilde{\cal P}_{\nu;s_1n_1,s_2n_2}^{(2);\beta\gamma}(w_0,w_3) &=&  \frac{[e\xi^\beta_\nu,\widetilde{\cal P}_{\nu
      }^{(1);\gamma}(w_3)]_{s_1n_1,s_2n_2}}{w_0-(s_1\epsilon_{n_1}-s_2\epsilon_{n_2})}\,,\label{eq:p2}\\
  \widetilde{\cal P}_{\nu;s_1n_1,s_2n_2}^{(3);\alpha\beta\gamma}(w,w_0,w_3) &=&  \frac{[e\xi^\alpha_\nu,\widetilde{\cal P}_{\nu
     }^{(2);\beta\gamma}(w_0,w_3)]_{s_1n_1,s_2n_2}}{w-(s_1\epsilon_{n_1}-s_2\epsilon_{n_2})}\,.\label{eq:p3}\quad~
\end{eqnarray}

The total current density is given as $\bm J(t) = \langle \hat{\bm
  J}(t)\rangle = J^{\alpha} \hat{\bm
  e}^{\overline{\alpha}}$ with
\begin{equation}
  J^{\alpha}(t) = -e{\cal
    D}\sum_{\tau s_1 s_2\atop n_1 n_2}v^{\alpha}_{\tau;s_2n_2,s_1n_1} \rho_{\tau;s_1n_1,s_2n_2}(t)\,.
\end{equation}
Correspondingly, the optical current can be expanded as $J^{\delta}(t)
= J^{(1);\delta}(t) + J^{(3);\delta}(t) + \cdots$ with
\begin{eqnarray}
  J^{(1);\tau}(t) &=& \int\frac{d\omega_3}{2\pi} {\sigma}^{(1);\tau\gamma}(\omega_3)
  E^{\gamma}({\omega_3}) e^{-i\omega_3 t} \,,\\
  J^{(3);\tau}(t)&=&  \int\frac{d\omega_1d\omega_2
    d\omega_3}{(2\pi)^3}
  \widetilde{\sigma}^{(3);\tau\alpha\beta\gamma}(w,w_0,w_3)
  E^{{\alpha}}({\omega_1})\notag\\
  && \times E^{{\beta}}({\omega_2})E^{{\gamma}}({\omega_3})
  e^{-i(\omega_1+\omega_2+\omega_3) t} \,,
\end{eqnarray}
where
\begin{eqnarray}
 \sigma^{(1);\tau\alpha}(\omega) &=& -e{\cal D}\sum_{\tau s_1 s_2 \atop n_1n_2} 
  v^\tau_{\nu;s_2n_2,s_1n_1}\notag\\
  && \times  \widetilde{\cal
    P}^{(1);\overline{\alpha}}_{\nu;s_1n_1,s_2n_2}(\hbar\omega+i\Gamma)
  \,,\\
  \widetilde{\sigma}^{(3);\tau\alpha\beta\gamma}(\omega_1,\omega_2,\omega_3)
  &=& -e{\cal D}\sum_{\tau s_1 s_2 \atop n_1n_2}
  v^\tau_{\nu;s_2n_2,s_1n_1}\notag\\
  && \times\widetilde{\cal
    P}^{(3);\overline{\alpha}\overline{\beta}\overline{\gamma}}_{\nu;s_1n_1,s_2n_2}(w,w_0,w_3)\,.
\end{eqnarray}
By substituting Eqs.~(\ref{eq:p1}-\ref{eq:p3}) we obtain
Eqs.~(\ref{eq:sigma1}) and (\ref{eq:sigma3}) in the main text. 

The independent Cartesian components of the third order conductivity
$\widetilde\sigma^{(3);dabc}(\omega_1,\omega_2,\omega_3)$ are expressed from
the circularly polarized components $\widetilde\sigma^{(3);\delta\alpha\beta\gamma}(\omega_1,\omega_2,\omega_3)$ as 
\begin{eqnarray}
  {}{\sigma}^{(3);xxyy} &=& {}{\sigma}^{(3);yyxx} \notag\\
    &=&
  \frac{1}{4}\sum_{\tau}\left({}{\sigma}^{(3);\tau\tau{\tau}\overline{\tau}}+
       {}{\sigma}^{(3);\tau{\tau}\overline{\tau}{\tau}}
    -
    {}{\sigma}^{(3);\tau\overline{\tau}{\tau}\tau}
    \right)\,,\notag\\
          {}{\sigma}^{(3);xyxy} &=& {}{\sigma}^{(3);yxyx}\notag \\
          &=&
  \frac{1}{4}\sum_{\tau}\left({}{\sigma}^{(3);\tau\tau{\tau}\overline{\tau}}-
       {}{\sigma}^{(3);\tau{\tau}\overline{\tau}{\tau}}
    +
    {}{\sigma}^{(3);\tau\overline{\tau}{\tau}\tau}
  \right)\,,\notag\\
        {}{\sigma}^{(3);xyyx} &=& {}{\sigma}^{(3);yxxy} \notag\\
        &=&
        \frac{1}{4}\sum_{\tau}\left(-{}{\sigma}^{(3);\tau\tau{\tau}\overline{\tau}}+
             {}{\sigma}^{(3);\tau{\tau}\overline{\tau}{\tau}}
    +
    {}{\sigma}^{(3);\tau\overline{\tau}{\tau}\tau}
    \right)\,,\notag
\end{eqnarray}
and
{\allowdisplaybreaks
\begin{eqnarray}
  {}{\sigma}^{(3);xyxx} &=& - {}{\sigma}^{(3);yxyy}\notag\\
        &=&
  \frac{i}{4}\sum_{\tau}\tau\left({}{\sigma}^{(3);\tau\tau{\tau}\overline{\tau}}+
    {}{\sigma}^{(3);\tau{\tau}\overline{\tau}{\tau}}
    -
    {}{\sigma}^{(3);\tau\overline{\tau}{\tau}\tau}
    \right)\,,\notag\\
          {}{\sigma}^{(3);xxyx} &=&- {}{\sigma}^{(3);yyxy} \notag\\
          &=&
  \frac{i}{4}\sum_{\tau}\tau\left({}{\sigma}^{(3);\tau\tau{\tau}\overline{\tau}}-
    {}{\sigma}^{(3);\tau{\tau}\overline{\tau}{\tau}}
    +
    {}{\sigma}^{(3);\tau\overline{\tau}{\tau}\tau}
    \right)\,,\notag\\
          {}{\sigma}^{(3);xxxy} &=&- {}{\sigma}^{(3);yyyx} \notag\\
          &=&
  \frac{i}{4}\sum_{\tau}\tau\left(-{}{\sigma}^{(3);\tau\tau{\tau}\overline{\tau}}+
    {}{\sigma}^{(3);\tau{\tau}\overline{\tau}{\tau}}
    +
    {}{\sigma}^{(3);\tau\overline{\tau}{\tau}\tau}\right)\,.\notag
\end{eqnarray}}


%\bibliographystyle{apsrev4-1}
%\bibliography{/home/cheng/nutstoreA/REFERENCE/2D_nano.bib,/home/cheng/REFERENCE/BIB/optics,/home/cheng/REFERENCE/BIB/BIB.bib}
%\bibliography{/home/nutstore/REFERENCE/2D_nano.bib,/home/cheng/REFERENCE/BIB/optics,/home/cheng/REFERENCE/BIB/BIB.bib}

\end{document}